\documentclass[ twocolumn]{aastex62}
\hypersetup{linkcolor=red,citecolor=blue,filecolor=cyan,urlcolor=magenta}
\usepackage{filecontents}
\usepackage{booktabs}
\usepackage{amsmath}
\usepackage{amsfonts}
\usepackage{amssymb}
\usepackage{xcolor}

\newcommand{\re}{R$_e$}

\newcommand{\normre}{$R_e / \overline{R_e}$}
\newcommand{\normha}{L$_{H\alpha}$/$\overline{L_{H\alpha}}$}
\newcommand{\normuv}{L$_{FUV}$/$\overline{L_{FUV}}$}

\begin{document}
\font\myfont=cmr12 at 14pt
\title{\MakeUppercase {\large{ Testing the Relationship Between Bursty Star Formation and Size Fluctuations of Local Dwarf Galaxies }}}
\correspondingauthor{Najmeh Emami}
\email{nemami@umn.edu}
\author{ Najmeh Emami}
\affiliation{Minnesota Institute for Astrophysics, University of Minnesota, 116 Church St SE, Minneapolis, MN 55455, USA}
\affiliation{Department of Physics and Astronomy, University of California Riverside, Riverside, CA 92521, USA}
\author{ Brian Siana}
\affiliation{Department of Physics and Astronomy, University of California Riverside, Riverside, CA 92521, USA}
\author{Kareem El-Badry}
\affiliation{Department of Astronomy and Theoretical Astrophysics Center, University of California Berkeley, Berkeley, CA 94720, USA}
\author{David Cook}
\affiliation{Infrared Processing and Analysis Center, California Institute of Technology, Pasadena, CA, USA}
\author{Xiangcheng Ma}
\affiliation{Department of Astronomy and Theoretical Astrophysics Center, University of California Berkeley, Berkeley, CA 94720, USA}
\author{Daniel Weisz}
\affiliation{Department of Astronomy and Theoretical Astrophysics Center, University of California Berkeley, Berkeley, CA 94720, USA}
\author{Joobin Gharibshah}
\affiliation{Computer Science and Engineering Department, University of California,
Riverside,CA 92521, USA}
\author{Sara Alaee}
\affiliation{Computer Science and Engineering Department, University of California,
Riverside,CA 92521, USA}
\author{ Claudia Scarlata}
\affiliation{Minnesota Institute for Astrophysics, University of Minnesota, 116 Church St SE, Minneapolis, MN 55455, USA}
\author{Evan Skillman}
\affiliation{Minnesota Institute for Astrophysics, University of Minnesota, 116 Church St SE, Minneapolis, MN 55455, USA}

\begin{abstract}

Stellar feedback in dwarf galaxies plays a critical role in regulating star formation via galaxy-scale winds. Recent hydrodynamical zoom simulations of dwarf galaxies predict that the periodic outward flow of gas can change the gravitational potential sufficiently to cause radial migration of stars. To test the effect of bursty star formation on stellar migration, we examine star formation observables and sizes of 86 local dwarf galaxies. We find a correlation between the R-band half-light radius (R$_e$) and far-UV luminosity (L$_{FUV}$) for stellar masses below 10$^8 $ M$_{\odot}$ and a weak correlation between the R$_e$ and H$\alpha$ luminosity (L$_{H\alpha}$). We produce mock observations of eight low-mass galaxies from the FIRE-2 cosmological simulations and measure the similarity of the time sequences of R$_e$ and a number of star formation indicators with different timescales. Major episodes of R$_e$ time sequence align very well with the major episodes of star formation, with a delay of $\sim$ 50 Myrs. This correlation decreases towards SFR indicators of shorter timescales
such that $R_e$ is weakly correlated with L$_{FUV}$ (10-100 Myr timescale) and is completely uncorrelated with L$_{H\alpha}$ (a few Myr timescale), in agreement with the observations. Our findings based on FIRE-2 suggest that the R-band size of a galaxy reacts to star formation variations on a $\sim50$ Myr timescale. With the advent of a new generation of large space telescopes (e.g., JWST), this effect can be examined explicitly in galaxies at higher redshifts where bursty star formation is more prominent.
\end{abstract}


\section{Introduction}
\label{sec:Intro}
Flat cosmological models with a mixture of dark energy, cold dark matter, and baryonic matter are consistent with the observations of structure formation on large-scales \citep[$>>$1 Mpc;][]{Bahcall_1999,Frenk_2012, Vogelsberger_2014, Schaye_2015, Planck_Collaboration_2020}. In this cold dark matter model ($\Lambda$CDM), dark matter is viewed as particles with weak self-interactions as well as negligible interactions with baryonic matter.
However, on galactic scales ($<$1 Mpc) there are several tensions between observations and predictions of $\Lambda$CDM \citep[For a comprehensive review see][]{Bullock_2017}. 

Dwarf galaxies, with stellar masses $\lesssim 10^9 M_{\odot}$, are known to challenge CDM predictions on small scales \citep{brook_2019}. It has long been recognized that the observed dark matter halos of dwarf galaxies exhibit a different core density profile \citep{Flores_1994, Moore_1994} than the predicted dark matter-only Navarro–Frenk–White (NFW) \citep{Navarro_1997} profile.

The NFW density profile rises steeply at small radii \citep{Navarro_2010}. However observations of rotation curves of dwarf galaxies reveal rather a constant-density profile in apparent contrast with the NFW profile \citep{McGaugh_2001, Marchesini_2002, Simon_2005, Kuzio_2008, Oh_2015, Relatores_2019}, referred to as the "cusp-core" problem. Two solutions have been proposed in the literature.

The first solution is a modification to the $\Lambda$CDM in which dark matter particles are posited to be strongly self-interacting \citep{Spergel_2000, Kaplinghat_2016}. In the self-interacting dark matter (SIDM) model, dark matter collisions allow energy exchange between particles and thermalize the inner halo over the cosmological timescale, leading to a cored central density \citep{Vogelsberger_2012, Peter_2013, Elbert_2015, Fry_2015, Sameie_2020}.

The second solution to the cusp-core problem is to implement baryonic feedback \citep[in forms of supernova explosions, radiation pressure, photoelectric heating, photoionization and stellar winds;][]{Navarro_1996, Governato_2010, Sales_2014, Hopkins_2020}. In these models, outflows that are driven by baryonic feedback over the course of star formation, are capable of inducing fluctuations in the central gravitational potential of the galaxy, transferring energy into dark matter particles, and ultimately flattening the dark matter core density profile in dwarf galaxies \citep{Governato_2010, Sales_2010, Pontzen_2012, Chan_2015, Tollet_2016, Read_2016}. This alternative solution requires high-resolution hydrodynamical simulations to model the inhomogeneous inter-stellar medium (ISM) and couple the star formation to high density gas regions \citep{Hopkins_2014, Hopkins_2018}. Implementation of baryonic feedback into CDM-only simulations has shown to further explain several other characteristics of dwarf galaxies such as their low stellar-to-halo mass ratio \citep{Hirschmann_2012, Moster_2013, Behroozi_2013}.

Other than forming a cored dark matter density profile, baryonic feedback is predicted to have additional impacts on the dynamical and morphological properties of dwarf galaxies \citep{Elbadry_2016, el-badry_2017, Read_2019}, many of which are shown to be in agreement with recent observations. For instance, outflows caused by supernova explosions result in bursty star formation rates in dwarf galaxies \citep{McQuinn_2010, Sparre_2017}, which is consistent with the different star formation rates (SFR) inferred from SFR indicators that probe different time scales in local and high-redshift dwarf galaxies \citep{Weisz_2012, dominguez_2015, Guo_2016,Mehta_2017, Emami_2019, caplar_2019,Faisst_2019, flores_2020}. Outflows can also explain the lack of coherent disc formation in most irregular dwarf galaxies \citep{Wheeler_2017} as well as an observed specific SFR- gas velocity dispersion (sSFR-$\sigma$) correlation in this mass range \citep{Stilp_2013, Cicone_2016, hirtenstein_2019, pelliccia_2020}. Furthermore, outflowing winds that are powered by episodic bursts of star formation drive radial migration of star and gas particles over the course of star formation \citep{ DeYoung_1994, Graus_2019, Mercado_2020}, which is consistent with metallicity and inverted age gradients in low-mass galaxies \citep{Aparicio_2000, Stinson_2009, Hidalgo_2013, Vargas_2014, McQuinn_2017, Jafariyazani_2019}.

The predicted radial migration has another consequence; the outflowing winds can also impact the size of the galaxy. In particular, during bursts of star formation gas and stars move into the center of the galaxy until the first supernova explodes. Subsequent SNe explosions will yield a shallow gravitational potential, gas and stellar migration towards larger radii resulting in more extended galaxy morphologies.
Over time, the expelled gas will cool and collapse back onto the galaxy center triggering the next burst of star formation.
This causes size fluctuations during bursts of star formation across the galaxy \citep{Elbadry_2016}.
Yet this prediction has not been observationally confirmed for the real dwarf galaxies.

In this work, we aim to test this prediction for a sample of local dwarf galaxies with a stellar mass range of $10^{7}- 10^{9.6} M_{\odot}$, the same regime suggested by \citet{Elbadry_2016}. We will study the relationship between bursty star formation and size fluctuations in both observed and simulated galaxies and provide physical intuitions for our findings.

An overview of this paper is as follows.
We describe the observed sample and our measurements of the galaxies' stellar mass, dust corrected luminosities, and R-band size in Section \ref{sec:data}. In Section \ref{sec:results} we discuss our observed findings. In Section \ref{sec:simulation} we analyze a suite of simulated dwarf galaxies, analogous to our observed sample, from FIRE-2 simulations and provide physical interpretations about the effect of bursty star formation on the galaxy's size fluctuation. We discuss previous works and the implications for the era of James Webb Space Telescope in Section \ref{sec:discussion} and lastly summarize our results in Section \ref{sec:summary}.

Throughout this work we assume that the default $\Lambda$CDM cosmology
has parameters H$_0$=67.4 km $s^{-1}$ Mpc$^{-1}$, $\Omega_{m}$ = 0.315, $\Omega_{\Lambda}$= 0.685 \citep{Planck_Collaboration_2020}


\section{Observations \& Measurements}
\label{sec:data}

In this work, we use a sample of galaxies from the Local Volume Legacy (LVL) Survey \citep{Kennicutt_2008}. LVL is an unbiased and statistically complete sample of 258 nearby field star-forming galaxies. Galaxies in the LVL sample are selected based upon a set of criteria including galaxies that are within 11 Mpc ($D\leq11$), lie outside the galactic plane ($|b| > 20^{\circ}$), are brighter than $B = 15$ mag, and span an RC3 Type galaxy range T$\geq 0$ (i.e., galaxies of spiral and irregular morphologies later than S0a). Due to its volume-limited nature, the LVL sample is dominated by dwarf galaxies and is complete above $10^{6.5}$ M$_{\odot}$ stellar mass.
The LVL public dataset consists of {\it GALEX} ultraviolet \citep{Lee_2011}, narrowband H$\alpha$ imaging \citep{Kennicutt_2008},  {\it Spitzer} IRAC and MIPS imaging \citep{Dale_2009}, optical MMT spectroscopy \citep{Berg_2012}, and optical ground-based imaging \citep{Cook_2014}.

About 70\% of the sample has deeper R-band imaging ($\sim$ 0.5 magnitude deeper than SDSS) \citep{Cook_2014} obtained from 1-2 meter telescopes located at Cerro Tololo Inter-American Observatory (CTIO), Kitt Peak National Observatory (KPNO), Vatican Advanced Technology Telescope (VATT), and the Wyoming Infra-Red Observatory (WIRO). The median pixel scale of the instruments are 0.5 arcsec pixel$^{-1}$ with a standard deviation of 0.1 arcsec pixel$^{-1}$.
For the other 30\% of the sample, we use SDSS r-band imaging taken from SDSS DR7 \citep[Sloan Digital Sky Survey;][]{abazajian_2009}. We use the data product produced in \cite{Cook_2014}. 

All the images are sky-subtracted and corrected for the contaminant sources, e.g., background galaxies and foreground stars.

The H$\alpha$ flux measurement is described extensively in \cite{Kennicutt_2008}. Different aperture configuration methods were applied on galaxies in the sample to ensure that the H$\alpha$ flux is integrated over the entire galaxy, not a subset.

The stellar masses and dust extinctions are obtained from \cite{Weisz_2012} and \citet{Johnson_2013}. Briefly, the masses are derived by fitting the observed UV, optical, and IR luminosities of each galaxy with the \citet{bruzual_2003} suite of stellar population synthesis models and a varying set of parameters (age, stellar metallicity, exponentially declining star formation histories, and dust). The H$\alpha$ and UV dust extinctions are derived from the ``energy balance" extinction correction method \citep{Kennicutt_2009, Hao_2011} by which the intrinsic luminosity is retrieved from the combination of unobscured (e.g., H$\alpha$ and UV) and obscured (FIR continuum) signatures of star formation. For the 30\% of the sample where FIR measurements are not available, a correlation-based attenuation correction is used \citep{Lee_2009}. For that, the H$\alpha$ dust attenuation is determined via a scaling relation between A$_{H\alpha}$ and B-band magnitude (M$_B$) \citep{Lee_2009}.
As shown in \cite{Lee_2009}, the scatter between A$_{H\alpha}$ and M$_B$ decreases with decreasing luminosity, such that for the relatively transparent dwarf galaxies which dominate our sample (-18$<$ M$_B$ $<$-14.7) the scatter is roughly 10-20\%.
The ${H\alpha}$ extinction corrections used here yield similar results to those used by \cite{Lee_2009}. The dust corrections derived by the application of energy balance and the empirical A$_{H\alpha}-$M$_{B}$ correlation are much smaller than the observed scatter in the H$\alpha$ luminosity and do not have a significant effect on our interpretations about the bursty star formation of the sample.
The UV dust attenuation ($A_{UV}$) is also estimated as the H${\alpha}$ attenuation (A$_{H\alpha}$) scaled by 1.8.
 
\begin{figure}
\begin{center}
\includegraphics[width=1\linewidth]{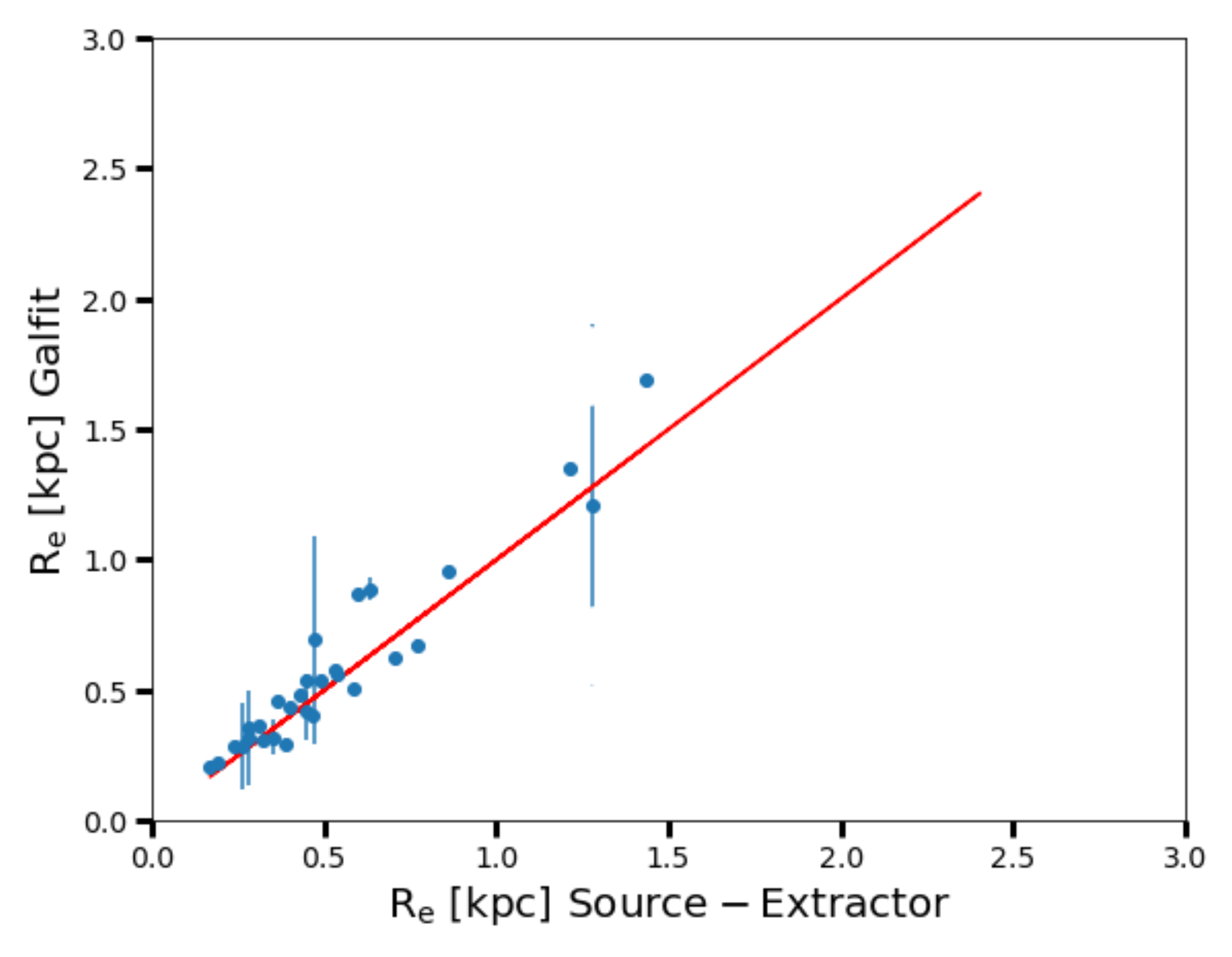}
\caption{A comparison between the half-light radii (\re) derived from Source-Extractor and GALFIT algorithms for a subsample of galaxies. The errors from GALFIT are very small with a median value of 2\%.}
\label{fig:galfit_vs_sextractor}
\end{center}
\end{figure}

We determine the sizes of the galaxies in the R-band images. This is because the galactic potential fluctuations caused by the outflowing winds impact the distribution of all stars of any age. Additionally, through the R band imaging, light from both young and old stellar populations can be assessed via the combination of strong nebular lines (i.e., H$\alpha$ which traces stars that are formed a few Myrs ago) and the stellar optical continuum (at a wavelength range of 5800-7300 \AA\ which probes stars that are formed over hundreds of Myrs). We run Source-Extractor \citep{Bertin_1996} to determine the half-light radii in the R-band. Since the images are sky-subtracted, we set the background value to zero in the code. For each galaxy, the zero point, pixel scale, and gain parameters are set by values stored in the header files. These parameters depend on the characteristics of the telescope with which the object was observed. We set the full-width at half maximum (FWHM) parameter to the seeing of the observation. 
We choose a flexible elliptical aperture to measure the flux of the detected objects inside that aperture as described in \cite{Kron_1980}.
Whenever the image is highly resolved, which is mostly the case for objects that are closer than 3 Mpc, the Source-Extractor algorithm segments the galaxy into multiple sources. In these cases we smooth the image by applying a Gaussian kernel with a minimum FWHM necessary for the Source-Extractor to recognize the galaxy as a single source. 
It may be of concern that the smoothing procedure washes out the galaxies' substructures, such as the spiral arms, therefore the measured half-light radii are smaller than when the smoothing is not applied. However, the convolution that we perform is with a much (factor of $\sim$ 5) smaller profile than the ultimate sizes measured in these galaxies. This means that the measured sizes will not be significantly increased by performing this convolution. 
Source-Extractor yields the half-light radius from a circular photometric aperture within which 50\% of the total flux of the galaxy is enclosed. The choice of circular aperture is an acceptable approximation since the galaxies are mostly circular or irregular (few edge-on disks). For a sanity check, we also run GALFIT \citep{Peng_2002} for galaxies in the sample. We fit
our galaxies with the Sersic profiles and a free Sersic index. Giving the Source-Extractor parameters as an initial guesses to GALFIT, we get half-light radii that are consistent with those from Source-Extractor. We compare the half-light radii derived from Source-Extractor and GALFIT for a subsample of our galaxies in Figure \ref{fig:galfit_vs_sextractor}. The errors derived from the GALFIT algorithm are very small with a median value of 2\%. 

To be consistent with \citet{Elbadry_2016}, we only keep galaxies with axis ratios $>0.5$ and stellar masses between $10^{7}- 10^{9.6} M_{\odot}$ where burstiness is predicted to be more prominent compared to other mass ranges \citep{Elbadry_2016}. This leaves us with a final sample of 86 galaxies.

\section{Results}
\label{sec:results}

We present our size measurements as a function of stellar mass in Figure \ref{fig:lvl_size_mass}. The logarithm of half-light radius (\re) in our LVL sample linearly increases with the logarithm of mass with a slope of $0.282 \pm 0.025$. Our \re\ values are in agreement with the r-band effective radii of SDSS z$\sim$0 galaxies measured in the NASA SLOAN Atlas \citep{blanton_2011}. 

At fixed mass, there is $\sim$ 0.8 dex scatter inferred by the standard deviation.
This scatter is the main driver of this work and is predicted to be partially driven by the feedback-driven outflows generated by episodic bursts of star formation \citep{Elbadry_2016}.  Here we aim to examine this prediction and see whether there is any relationship between the size fluctuation and star formation change in the observed low-mass galaxies. 

\subsection{Observed Size-Luminosity Relation}
\begin{figure}
\begin{center}
\includegraphics[width=1\linewidth]{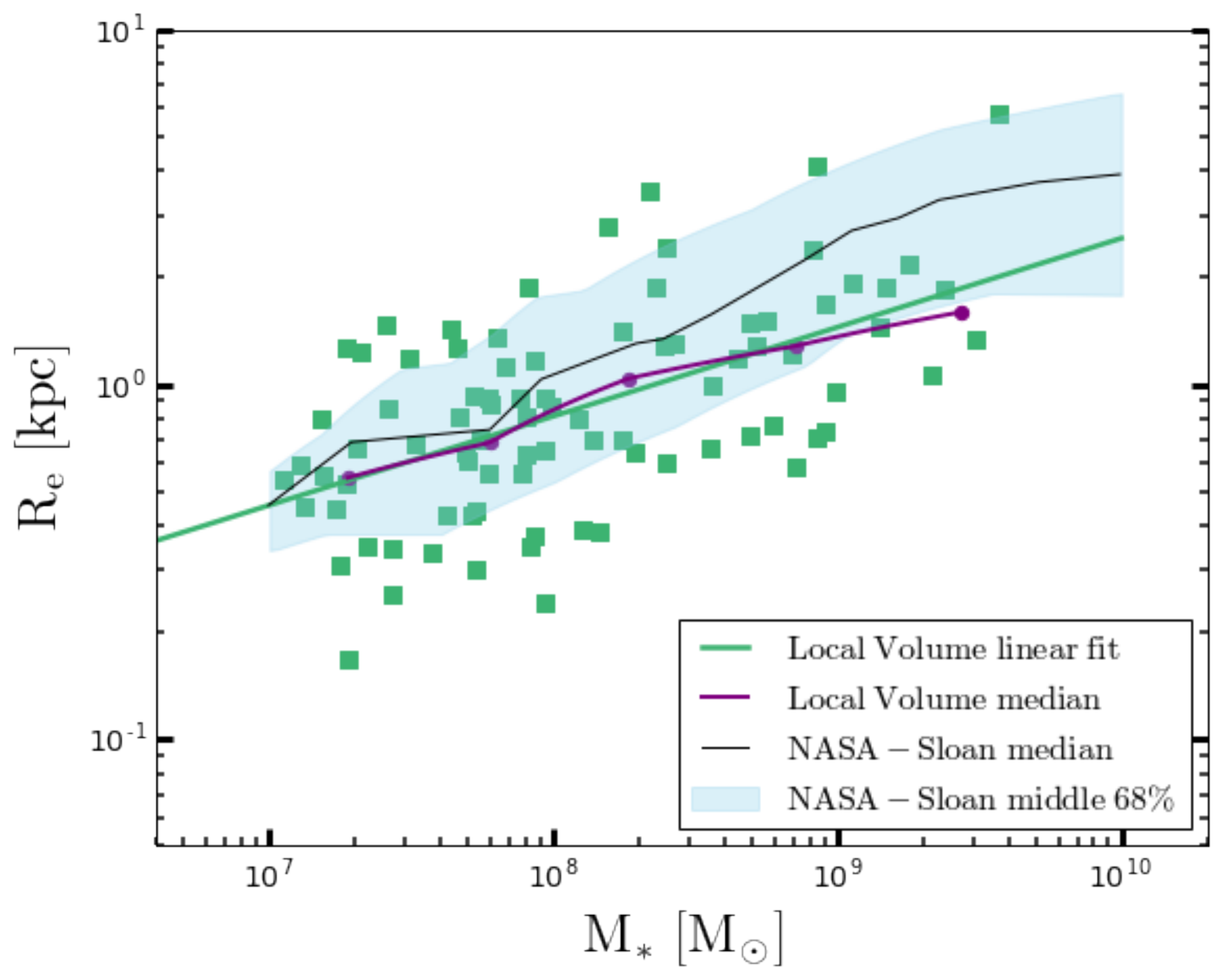}
\caption{R-band half-light radius (\re) as a function of stellar mass for the LVL sample (green points). There is a correlation between the two quantities in the log scale with a slope of 0.282 $\pm$ 0.025 shown as green line. The purple points are the median of the LVL sample for five mass bins. The typical \re\ errors for the individual galaxies are about 2\% . The blue shaded area and the black curve are the median and 1$\sigma$ scatter of the NASA-Sloan Atlas at z$\sim$0 \citep{blanton_2011}. The median of the NASA-Sloan sample is very close to the median and the linear fit to our LVL sample. The standard deviation is $\sim$ 0.8 dex in \re\ at $10^8 M_{\odot}$ stellar mass. }
\label{fig:lvl_size_mass}
\end{center}
\end{figure}

In order to test the relation between the size and the star formation rate (o equivalently, the luminosity) of the sample, we need to first subtract off the mass dependency from the \re\ distribution. In particular, at any given mass, we are interested in each galaxy's deviation in \re\ from the average trend shown in Figure \ref{fig:lvl_size_mass}. For this, we introduce a new definition of size which is the ratio of \re\ to the average \re\ (\normre). Technically, at each point we divide the observed \re\ by the average $\overline{R_e}$ derived from the line fit through the log(\re)-log(M$_*$) relation and refer to it normalized-to-average size.

\begin{figure}
\begin{center}
\includegraphics[width=1\linewidth]{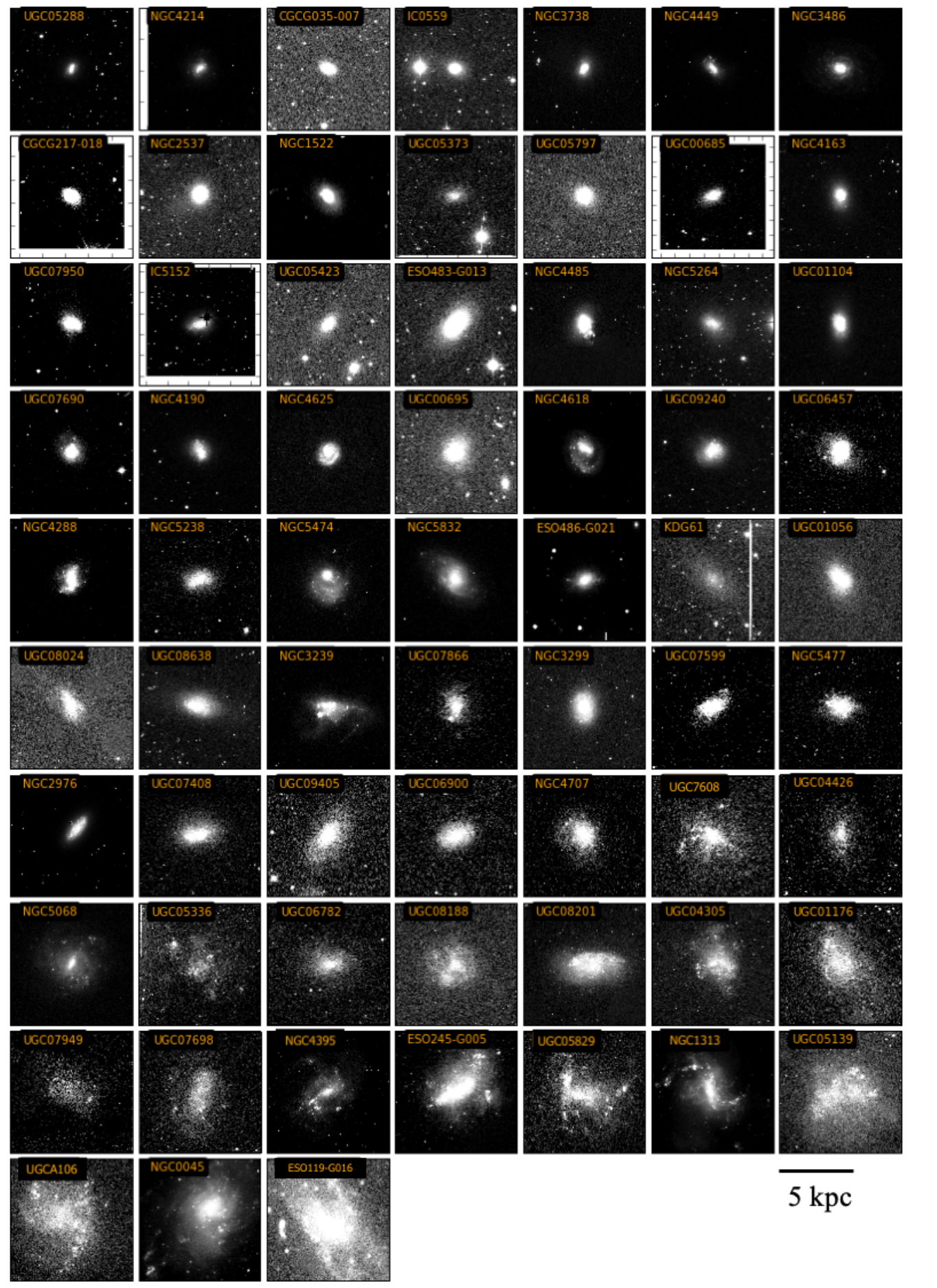}
\caption{Postage stamps of R-band images of the LVL sample which are scaled to follow the fit relation shown in Figure \ref{fig:lvl_size_mass} and are also scaled by their distance and are sorted by their \normre\ (half-light radius relative to the mean). Each stamp is labeled by the galaxies' LVL ID name.}
\label{fig:galaxies_image}
\end{center}
\end{figure}

Figure \ref{fig:galaxies_image} shows postage stamps of R-band images of a sample of LVL galaxies which are scaled by their distances as if they are all at the same distance and then are sorted by their \normre. The top left image shows the most compact galaxy in the sample with the smallest \normre\ while the image at the bottom-right indicates the galaxy with the largest \normre.

Next, we calculate the dust-corrected H$\alpha$ and far-UV (FUV) luminosities of the sample \citep{Weisz_2012}. H$\alpha$ and FUV luminosities are the two common probes of the star formation rates in galaxies with timescales of 5 and 10-100 Myrs respectively, depending upon the star formation systems \citep{kennicutt_1998, Emami_2019, flores_2020}.

 In order to take out the mass dependency from the H$\alpha$ (and FUV) luminosity distributions, we divide the H$\alpha$ (FUV) luminosity at a given mass by the average luminosity derived from the line fit through the log L$_{H\alpha}$ (log L$_{UV}$) and log M$_{*}$ relation at that mass, the same procedure we used to derive the normalized-to-average size values (\normre).

We then split the sample into four mass bins with equal number of galaxies in each bin and investigate the size-luminosity relation in each mass bin assuming that galaxies of similar stellar masses likely share common star formation properties \citep{Weisz_2012,Emami_2019}. In particular, lower mass galaxies should have a lower potential well and thus may be more likely to show signs of episodic SF in their sizes.

\begin{figure*}
\includegraphics[width=0.49\linewidth]{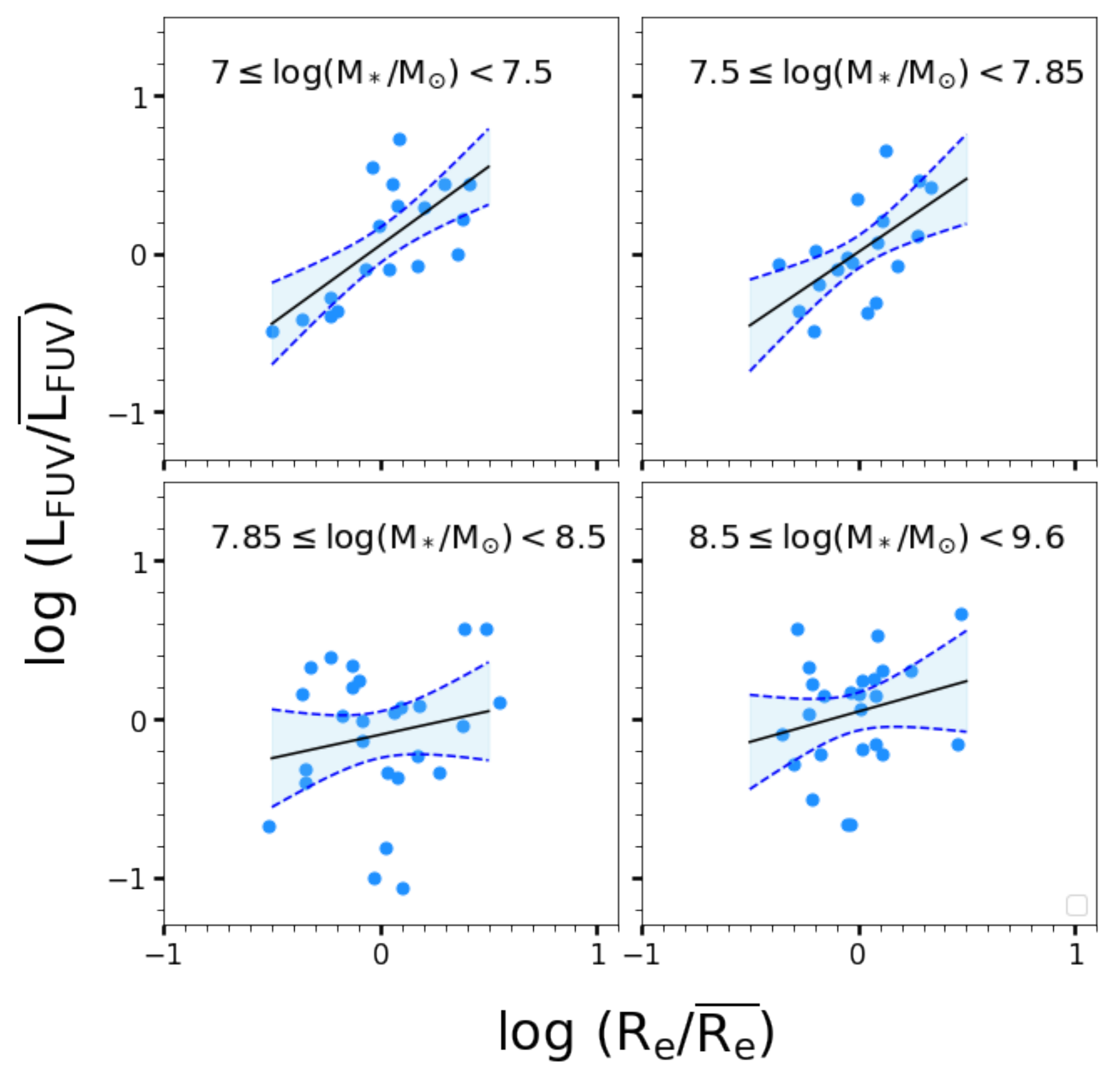}
\includegraphics[width=0.49\linewidth]{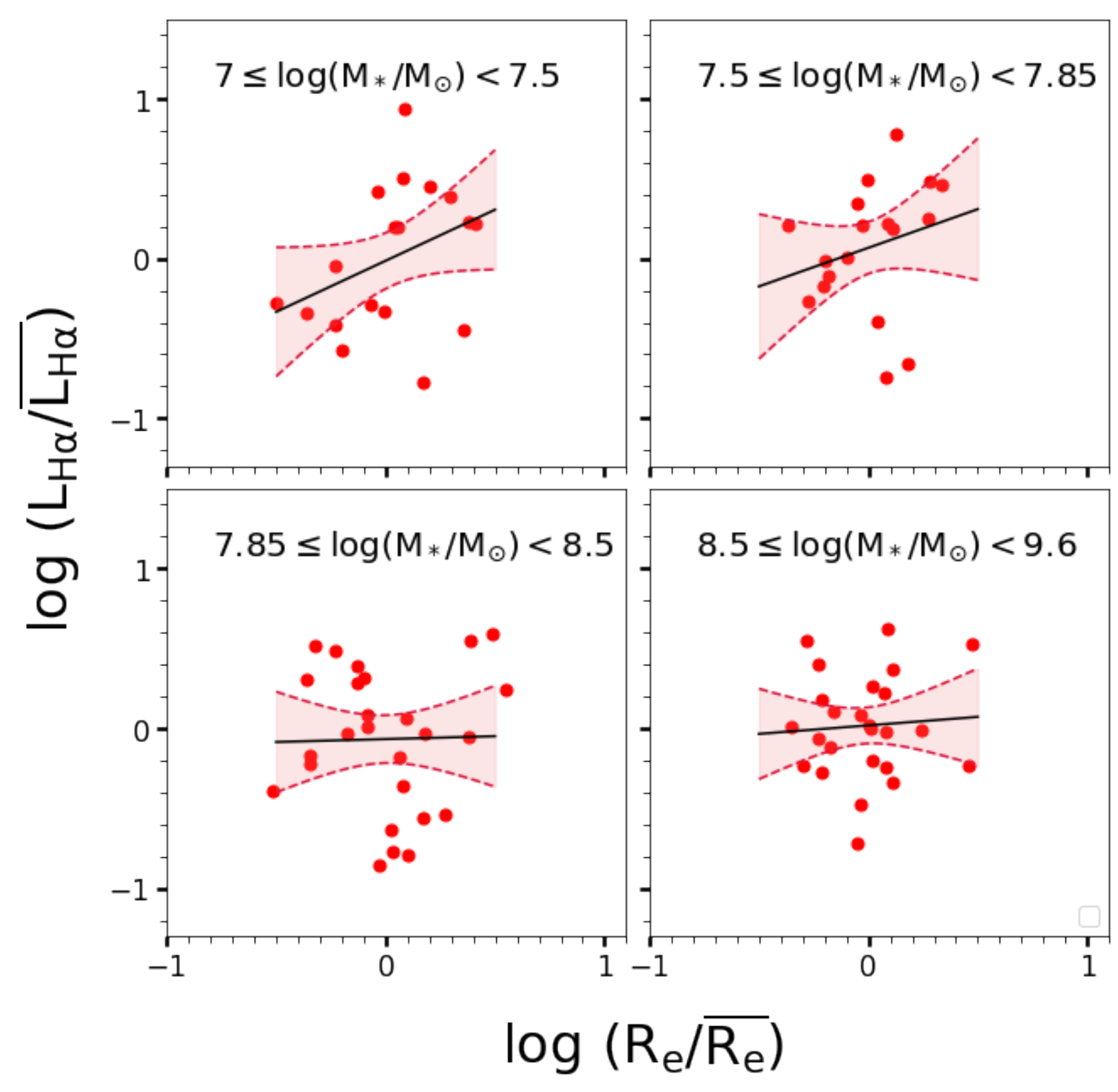}
\caption{Left: Logarithm of normalized-to-average far-UV luminosity density (\normuv) vs. logarithm of normalized-to-average half-light radius (\normre). A linear fit and 1$\sigma$ scatter is shown as black lines and blue shaded areas for each mass bin. There is a clear correlation at $10^{7} \leq$ M$_{*} < 10^{7.85}$ M$_{\odot}$ mass range which becomes weaker at larger masses. This can be interpreted as bursty star formation causes a size fluctuation in low-mass galaxies and that this size fluctuation reacts to the star formation change over a timescale similar to that of the FUV luminosity. Right: Logarithm of normalized-to average ${H\alpha}$ luminosity (\normha) vs. logarithm of (\normre). There is a positive slope with large scatter at masses below $10^{7.85}$ M$_{\odot}$ which becomes shallower towards larger masses.}
\label{fig:lvl_re_sfh}
\end{figure*}

Figure \ref{fig:lvl_re_sfh} shows the logarithm of normalized-to-average H$\alpha$ and FUV luminosities, namely \normha\ and \normuv\, as a function of log(\normre) for four mass bins. We find a positive correlation between log(\normuv) and log (\normre) at masses $<10^{7.85} M_{\odot}$. The correlation becomes weaker and more scattered towards larger mass bins. For the log(\normha)-log(\normre) relation, we also see a positive slope but less steep and more scattered than the log(\normuv)-log (\normre). We report the slopes of the best fit relationships for each mass bin as well as the Spearmann-rank correlation coefficients in Table \ref{table:lvl_size_sfr_slope}. A larger Spearmann-rank correlation coefficient indicates a stronger correlation between two quantities.

\begin{table*}
    \centering
    \begin{tabular}{c|c|c}
        \hline
       $Log(M_*/M_{\odot}$)  & Log(\normuv)-log(\normre)  & Log(\normha)-log(\normre) \\
       
         & slope ($\rho$)   & slope ($\rho$)\\
       \hline
       \hline
        7-7.5  & 1.00+/-0.28  (0.66)  &  0.70+/-0.40   (0.37)\\
        7.5-7.85 & 0.91+/-0.31  (0.65)  &  0.50+/-0.50   (0.37)\\
        7.85-8.5 & 0.26+/-0.32  (0.03)  &  0.03+/-0.33  (-0.06)\\
        8.5-9.6  & 0.33+/-0.34  (0.24)  &  0.09+/-0.32  (0.05)\\
        \hline
    \end{tabular}
    \caption{The slopes and Spearmann-rank correlations ($\rho$) of the Logarithm of far-UV and ${H\alpha}$ normalized-to-average luminosities vs. logarithm of normalized-to-average half-light radius (\normre) for four stellar mass bins in the LVL sample.}
    \label{table:lvl_size_sfr_slope}
\end{table*}

In order to better understand these observed trends, we compare to dwarf galaxies in hydrodynamical zoom-in simulations

\section{Hydrodynamical Simulations}
\label{sec:simulation}

In Section \ref{sec:results}, we found that for masses below $10^{7.85}$ M$_{\odot}$, the normalized-to-average size is correlated with the FUV normalized-to-average luminosity at fixed stellar mass. 
This supports predictions from hydrodynamical simulations \citep{Elbadry_2016} suggesting that the size fluctuation and stellar migration in low mass systems are consequences of feedback-driven outflows which in turn are driven by the cumulative effects of many starburst episodes over time. 

Based on this argument, star formation is triggered when cold gas accretes into the center of the galaxy and forms stars. Stellar feedback then creates a galactic wind and drives gas to larger radii, resulting in a reduction in star formation. Since the velocity of the expelled gas does not exceed the escape velocity \citep{muratov_2015} it soon cools, re-accretes back to the center and resumes star formation. This periodic gas displacement across the galaxy creates fluctuations in the central gravitational potential which in turn leads to variation in the stellar distribution and the measured size of the galaxy.
This theoretical picture can explain the relationship between the FUV luminosity and size in our observed low-mass sample. 

However, it is not yet understood why this dependency is less effective between the size fluctuation and short timescale SFR indicators, i.e., H$\alpha$ luminosity. Since H$\alpha$ and FUV luminosities trace the star formation rate on different time scales \citep{flores_2020}, a detailed investigation on the time scale over which the galaxy size fluctuation and star formation variation occur is critical.
In this section we aim to compare the SFHs with the half-light radius time sequences for a set of simulated galaxies in order to find quantitative explanations for our observed findings.

Feedback in Realistic Environments \citep[FIRE-1, FIRE-2;][]{Hopkins_2014, Hopkins_2018} is a set of simulations with models for star formation and stellar feedback implemented within cosmological baryonic simulations. In the FIRE model, energy and momentum inputs from stars are taken from stellar evolution calculations and models of the unresolved Sedov-Taylor phase, without tunable parameters. The simulations have been successful in reproducing several key observables, including
realistic galactic outflows \citep{muratov_2015, Muratov_2017}, the dense HI
content of galaxy haloes \citep{Faucher-Gigu_2015}, the mass-metallicity relation \citep{Ma_2016}, the mass-size relation and
age/metallicity gradients \citep{Elbadry_2016}, cored dark-matter
profiles \citep{Chan_2015, onorbe_2015}, stellar kinematics \citep{Wheeler_2017}, the Kennicutt-Schmidt relation \citep{Orr_2018}, observed abundance distributions \citep{Escala_2018}, and
a realistic population of satellites around MW-mass hosts \citep{Wetzel_2016}. 

Here, we use FIRE-2 simulations of eight isolated low-mass galaxies spanning over a mass range of $10^7$-$10^{9.6}$ M$_{\odot}$, analogous to our local observed sample, and explore their star formation and size evolution over the last 2 Gyrs (z $\sim 0.15-0$).

\subsection{Simulated Size-SFR relation}

\begin{figure*}
\begin{center}
\includegraphics[width=1\linewidth]{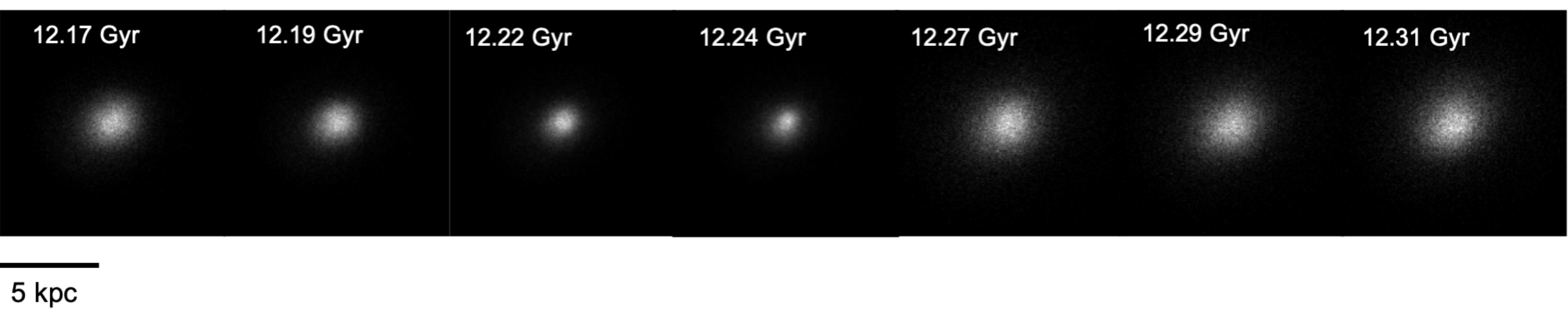}

\caption{Snapshots of R-band mock images of galaxy m11c in FIRE-2 simulations. It is evident that the simulated galaxy undergoes a significant size fluctuation across $\sim$150 Myrs.}
\label{fig:m11c_mock_images}
\end{center}
\end{figure*}

To calculate the R-band half-light radii of the simulated galaxies, we need to first generate mock R-band images of each snapshot in the simulation. The time spacing between consecutive snapshots is $\sim$ 20 Myrs. Following procedures in \citep{Chan_2018}, we generate a table of R-band luminosity-to-mass ratios of single stellar populations for a range of metallicity and age. For that we run the Flexible Stellar Population Synthesis \citep[FSPS;][]{Conroy_2009, Conroy_2010} model and assume the latest Padova stellar evolutionary tracks \citep{Marigo_2007, Marigo_2008}, Kroupa initial mass function \citep{Kroupa_2001}, and MILES stellar library. We then determine the R-band luminosity-to-mass ratio of each simulated star particle by interpolating the generated table at the particle's age and metallicity and then multiply the resulting luminosity-to-mass ratio by the mass of the star particle to get the R-band luminosity. We then project the R-band luminosities of each simulation output over a 20 kpc$\times$20 kpc region on to $1400^2$ mesh grids and generate mock R-band images. We have examined our analysis on a larger number of mesh grids and found our size measurements robust to spatial resolutions. Since most of these galaxies are spherical, the size measurements are independent from the orientation we choose to view the galaxy. We show snapshots of R-band mock images of galaxy m11c in Figure \ref{fig:m11c_mock_images} which indicates the galaxy undergoes a significant size fluctuation across $\sim$150 Myrs.

\begin{figure}
\begin{center}
\includegraphics[width=1\linewidth]{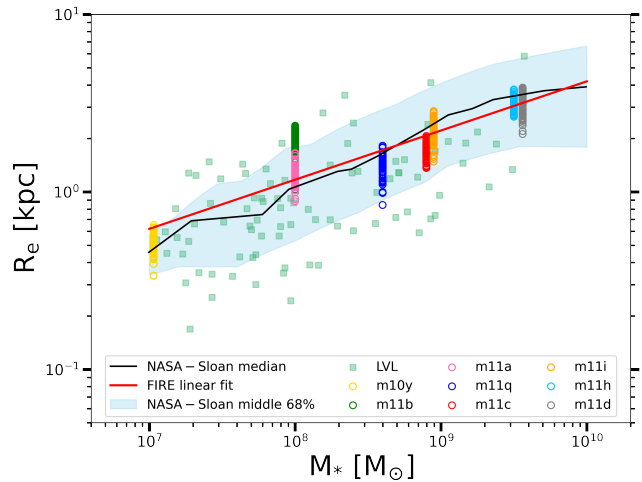}

\caption{R-band half-light radius (\re) as a function of stellar mass in FIRE-2 galaxies. \re\ is measured in 100 snapshots during the last two Gyrs. The red line is the linear fit to the FIRE-2 data points.
The size measured for the FIRE-2 galaxies are consistent with those of the observed LVL galaxies (green squares).
The blue shaded area and the black curve are the median and 1$\sigma$ scatter for the NASA-Sloan Atlas at z$\sim$0 \citep{blanton_2011}. The simulated galaxies are located within the 1$\sigma$ scatter in the NASA-Sloan sample and the median of the NASA-Sloan sample agrees very well with the FIRE-2 line fit.}
\label{fig:fire_reff_mass}
\end{center}
\end{figure}

We calculate the half-light radius for each snapshot in all simulated galaxies by passing the R-band mock images to Source-Extractor and use the same input parameters as those used for our observed galaxies (See Section \ref{sec:data}).
We also use a PSF equivalent to the median seeing of our observed sample (1.4 arcsec) and find that the chosen PSF is much smaller than the galaxy sizes.
In order to account for the potential impact of sky noise on our size measurements, we add stochastic sky backgrounds with an average surface brightness $\sim$ 26 mag arcsec$^{-2}$ to our mock images. We find that the differences in the estimated half-light radii with and without sky backgrounds are small and do not affect our analysis.
Figure \ref{fig:fire_reff_mass} shows the R-band half-light radius as a function of stellar mass for FIRE-2 galaxies. 
The distribution of the simulated galaxies is consistent with that of the observed LVL galaxies in the \re-Mass parameter space.
 Also, similar to our observed sample, we find our size measurements in agreement with those of z$\sim$0 SDSS galaxies \citep{blanton_2011}. We also see that \re\ fluctuates by more than a factor of 2 over $\sim$ 200 Myr in most of the galaxies.

\subsection{Size fluctuation as a consequences of bursty star formation}

\begin{figure*}
\begin{center}
\includegraphics[width=1\linewidth]{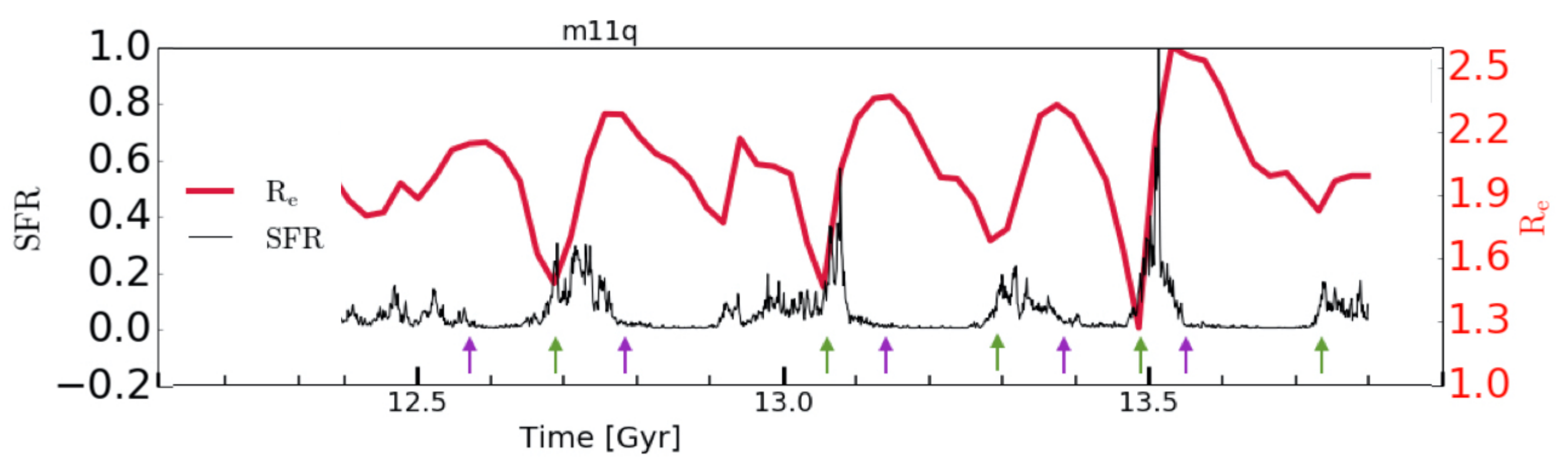}

\caption{Half-light radius (\re) time sequence (red) and SFH (black) for galaxy m11q. Starbursts with time durations of 100-200 Myr are evident in the SFH, consistent with the durations of star bursts observed in the local low-mass galaxies \citep{McQuinn_2012} . There is a delay in the \re\ to response to the star formation rate variations. \re\ is at its lowest when SFR has just begun rising (green arrows) and is at its highest once the SFR shuts down (purple arrows). It can be inferred that episodic bursts of star formation power energetic outflowing winds that drive away gas, shallow the gravitational potential, and cause the stars to move outward. This causes the galaxy's size to fluctuate dramatically on few 100 Myr timescales.}
\label{fig:m11q_Reff_sfh}
\end{center}
\end{figure*}

Next, we compare the SFHs with the \re\ time sequences of the simulated galaxies. Figure \ref{fig:m11q_Reff_sfh} shows the SFH and \re\ time sequences for galaxy m11q ($10^{8.6} M_{\odot}$). The local minima and maxima in the \re\ time sequences coincide with certain phases in the SFH; \re\ is at its lowest when SF has just begun rising (nearly after a few Myrs, green arrows) and it peaks once SF shuts down (purple arrows). The physical interpretation is that when SF has just begun rising, gas in the center is highly concentrated and forms stars at small radii, and that the majority of the stellar mass is in the center. After supernovae begin to explode, the energetic outflowing winds drive away the gas, resulting in a shallow central gravitational potential which drives stars to outer radii.
The subsequent supernovae cumulatively drive gas and stars further away and the galaxy expands and reaches its maximum size. Once SF has completely shut off, the central potential returns to a relaxation phase and the cooled gas and stars reaccrete and the galaxy starts to contract until the next burst cycle begins. We see this phenomenon in all of our simulated galaxies at a mass range of $10^7$-$10^{9.6}$ M$_{\odot}$. Based on this argument and Figure \ref{fig:m11q_Reff_sfh}, it can be inferred that the size of the galaxy traces the SF changes with a {\it time delay}.

\subsection{What is the time delay from SF change to galaxy size change? }
In order to determine how long it takes the size of a galaxy to respond to the SF change, i.e. the time delay from star formation change to the galaxy size change, we calculate the cross-correlation between the \re\ time sequence and SFH for each galaxy in the FIRE-2 sample. Since our \re\ data points are spaced by 20 Myrs while our SFRs are sampled at every single Myr, it is very important to first smooth the time sequence of the latter to a time resolution close to that of \re\ and then sample the time sequences for 20 Myr intervals, otherwise it is likely that the long-term SF fluctuations will be lost behind the instantaneous SF fluctuations.

\begin{figure}
\begin{center}
\includegraphics[width=1\linewidth]{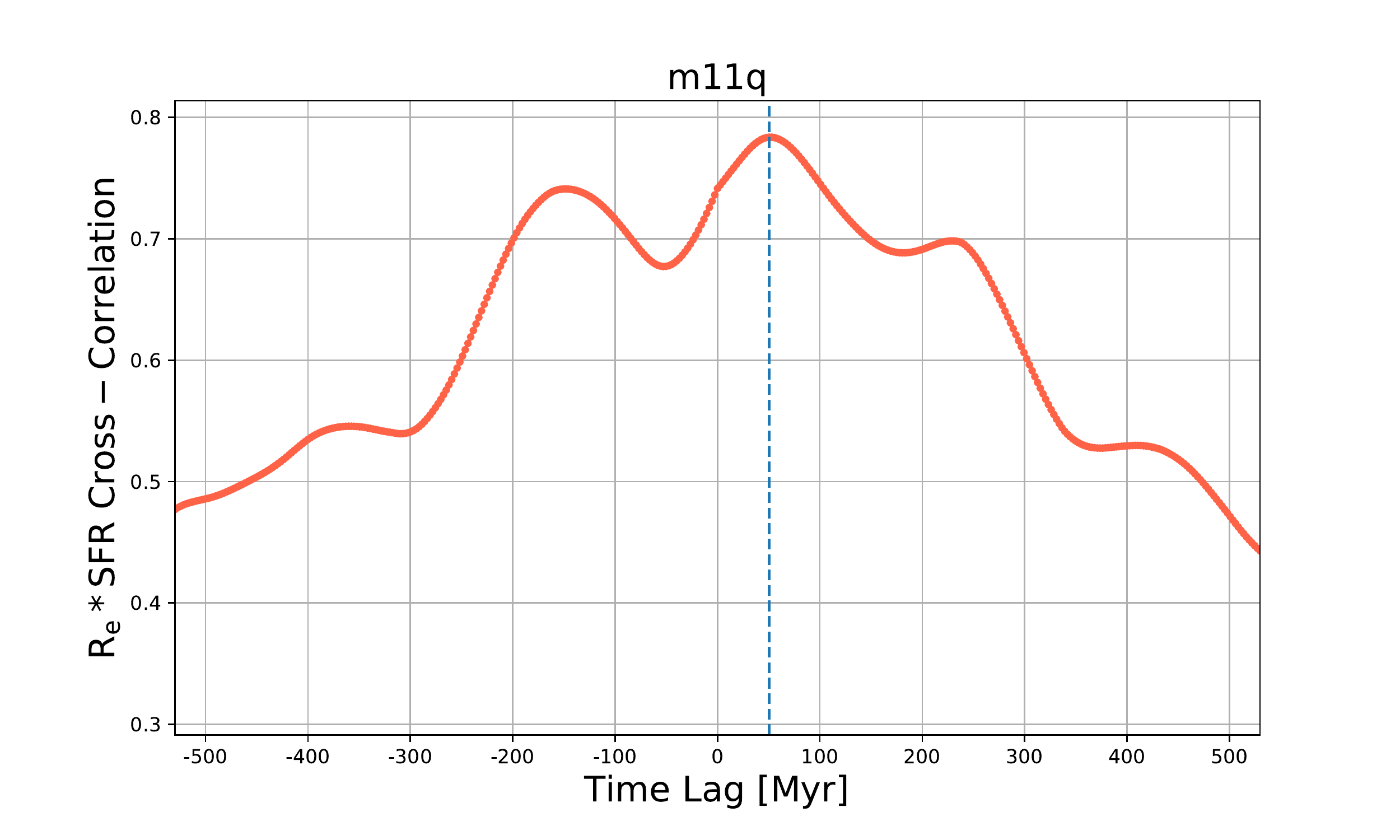}

\caption{Cross-correlation of half-light radius (\re) with SFR for galaxy m11q. The peak of the cross-correlation function at time lag=50 Myr suggests that the galaxy size traces the star formation change with a 50-Myr delay. The cross-correlation function for other FIRE-2 galaxies are provided in Appendix.}
\label{fig:cross_corr_m11q}
\end{center}
\end{figure}

In the cross-correlation algorithm, the correlation between \re\ and SFR time sequences with lag k is defined as:

\begin{equation}
   {\sum_{n} R_e[n+k].SFR^*[n]}
\end{equation}

\noindent where SFR$^*[n]$ is the complex conjugate of SFR at the {\it n}th snapshot and \re$[n+k]$ is the \re\ value at the ($n+k$)th snapshot. Figure \ref{fig:cross_corr_m11q} shows the cross-correlation between \re\ and SFR as a function of time lag for the galaxy, m11q. 
The peak in the cross-correlation plot indicates the time lag between the SFH and \re\ time sequences, by which the \re\ time sequences correlate best with the SFH if the SFH is shifted by that amount.
We show the same plots for the rest of the sample in Figure \ref{fig:cross_corr_all} in the Appendix. The time lag at the maximum cross-correlation is reported in Table \ref{table:cross_corr}. In six out of eight galaxies, the cross-correlation peaks at $\sim$ 50 Myr time lag. This implies that it takes the galaxy's size and the stellar migrations roughly 50 Myrs to react to any changes in the star formation rate. For the other two galaxies (m11b, m11h), the  cross-correlation peaks at zero Myr time lag between \re\ and SFR. This is because in these two galaxies the star bursts occur immediately one after another such that there is no quenching time between consecutive bursts, making it difficult for the cross-correlation algorithm to identify individual extrema in the SFH and \re\ time sequences. 

We also perform the cross-correlation algorithm on three SFR observables with different timescale sensitivities, including the H$\alpha$ luminosity, and far-UV (1500 \AA) and near-UV (3500 \AA) luminosity densities. Our goal is to see which one of these SFR observables has the shortest time lag with the \re\ and thus, is better (positively) correlated with the galaxy size. The median of the time lags between \re\ and H$\alpha$, far-UV, and near-UV for all the galaxies are 42, 28, and 8 Myrs respectively. This means that \re\ and near-UV luminosity are more correlated with each other (compared to the \re\ with H$\alpha$ or far-UV) and near-UV luminosity traces the SFR on a comparable timescale to the \re\ changes. This is expected given that the far-UV and H$\alpha$ light come from short-lived, massive, and hot stars that trace the star formation change on a short time scale, while near-UV light comes from longer-lived, less massive stars that take a longer time to trace the star formation variations.

\begin{figure*}
\begin{center}
\includegraphics[width=1\linewidth]{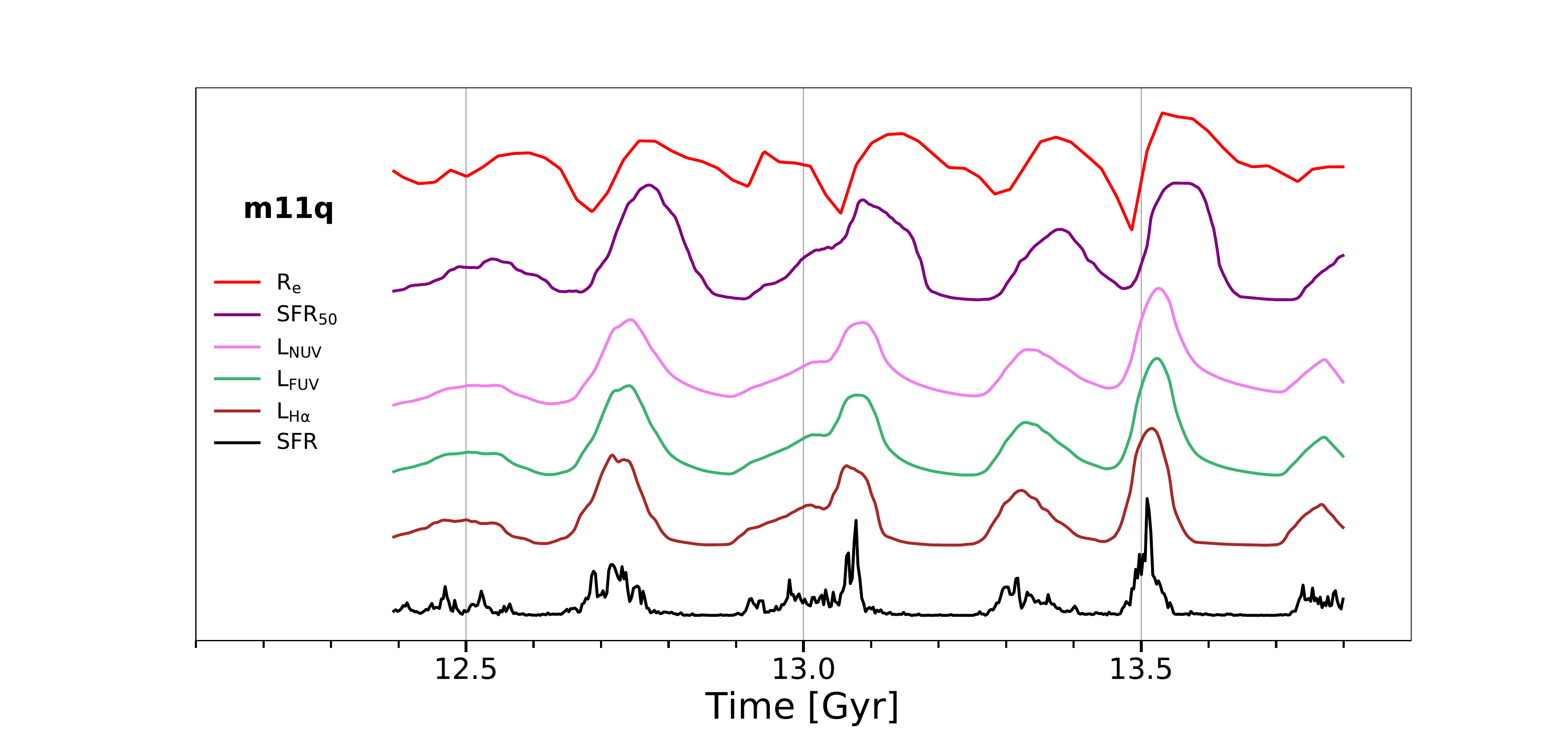}

\caption{Normalised half-light radius (\re) time sequences (red) is compared with the normalized time sequences of different SFR indicators in galaxy m11q. SFR indicators are sorted based on their correlation with the \re\ time sequence, with the highest correlation at the top (SFR$_{50}$, the SFR which is lagged by 50 Myr) and the lowest correlation at the bottom (L$_{H\alpha}$). Major episodes in \re\ are very well in phase with the major episodes in the SFR$_{50}$  time sequence and are completely out of phase with that of L$_{H\alpha}$. The SFH is also shown in black for reference. The complete figure set is available in the online journal for the rest of the FIRE-2 sample.}
\label{fig:m11q_timesequences}
\end{center}
\end{figure*}

In Figure \ref{fig:m11q_timesequences} we show the normalized \re\ time sequences along with the normalized SFH that lags by 50 Myr (SFR$_{50}$), as well as the normalized near-UV, far-UV, and H$\alpha$ time sequences for galaxy m11q. It is evident that a SFH with a 50 Myr time lag is well in phase with the \re\ time sequence. The correlation between \re\ and the three SFR observables decreases from near-UV, to far-UV, and to H$\alpha$ luminosity, respectively, until the \re\ and H$\alpha$ time sequences become fully uncorrelated. This can also be seen in almost all of the FIRE-2 galaxies (See online journal for the rest of the FIRE-2 sample).

In order to directly compare these results from simulated galaxies with our observed results in Section \ref{sec:results}, we plot all the SFR indicators as a function of \re\ in Figure \ref{fig:re_sfr_fire} and fit a line to each distribution. We present the slopes and the Spearmann-rank correlation coefficients ($\rho$) in Table \ref{table:re_sfr_fire}. It is evident that there is a positive slope in the \re-SFR$_{50}$ relation and the slope decreases from left to right, ultimately reaching an uncorrelated \re-L$_{H\alpha}$. We also see a shallow FUV-\re\ correlation in most of the galaxies, which is in agreement with our observed findings in Figure \ref{fig:lvl_re_sfh}. Unlike in the observations, however, we do see a correlation at all masses. This is in tension with what we found in our observed sample in which the correlation was stronger at masses below $10^{7.85} M_{\odot}$ and became less correlated towards higher masses. This is due to the bursty nature of SF in FIRE simulations in which the timescale of star formation bursts is much shorter than the timescale implied by observations of real galaxies at high mass \citep{muratov_2015,Sparre_2017, Emami_2019, Stern_2020}.
Furthermore, we see that in Figure \ref{fig:re_sfr_fire}, for all the four SFR indicators (SFR$_{50}$, L$_{NUV}$, L$_{FUV}$, and L$_{H\alpha}$), the slopes are biased by a few points that show very low SFR. These points are associated with the post-burst phases that are followed by long periods of no star formation. In these cases, SFR decreases rapidly, while \re\ takes a longer time to decrease and reach an equilibrium. One example can be seen in Figure \ref{fig:m11q_timesequences} for which during the quenching time at 13.55 Gyr and the subsequent burst at 13.75 Gyr, SFR decreases rapidly while \re\ decreases at a much lower rate.

\begin{table*}[]
    \centering
    \begin{tabular}{c|c|c}
        \hline
       Name & Log(M$_*$/M$_{\odot}$)  & \re-SFR Time Lag [Myr] \\ & z $\sim$ 0 & (Cross-Correlation) \\
       \hline
      \hline
        m10y & 7.03 & 48  \\
        m11a & 8.0 & 52 \\
        m11b & 8.0 & 0  \\
        m11q & 8.6 & 50 \\
        m11c & 8.9 & 63 \\
        m11i & 8.95 & 43 \\
        m11h & 9.5 & 0 \\
        m11d & 9.56 & 56\\
        
        \hline
    \end{tabular}
    \caption{ Time lag at the maximum cross-correlation between SFR and half-light radius (\re) in FIRE-2 galaxies.}
    \label{table:cross_corr}
\end{table*}


\begin{figure*}
\begin{center}
\includegraphics[width=1\linewidth]{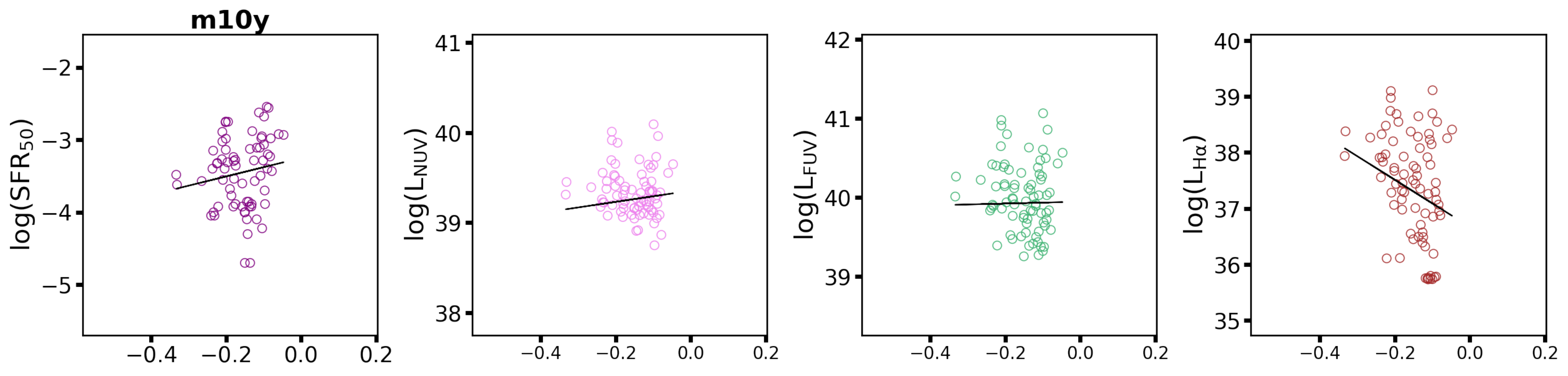}
\includegraphics[width=1\linewidth]{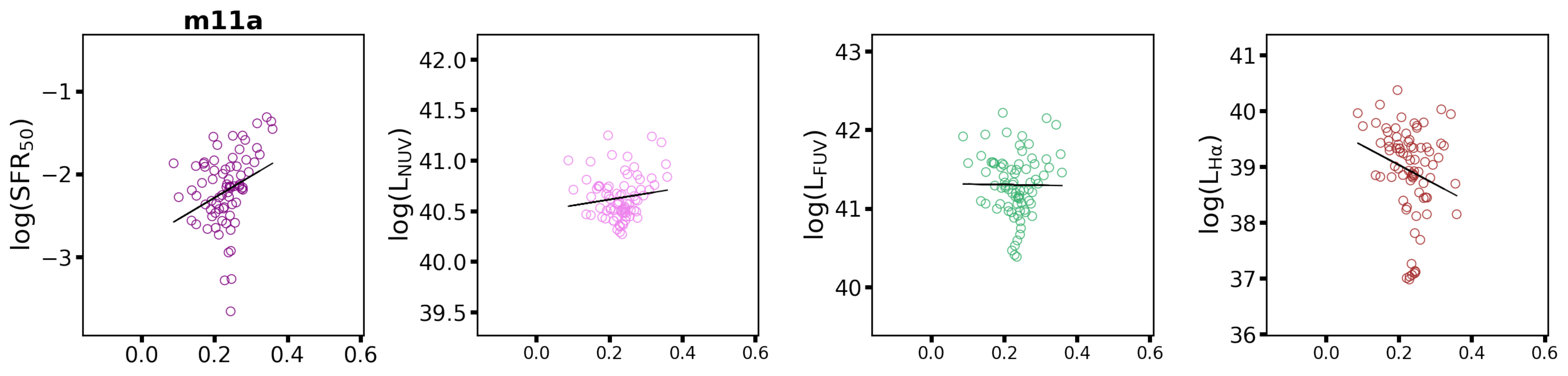}
\includegraphics[width=1\linewidth]{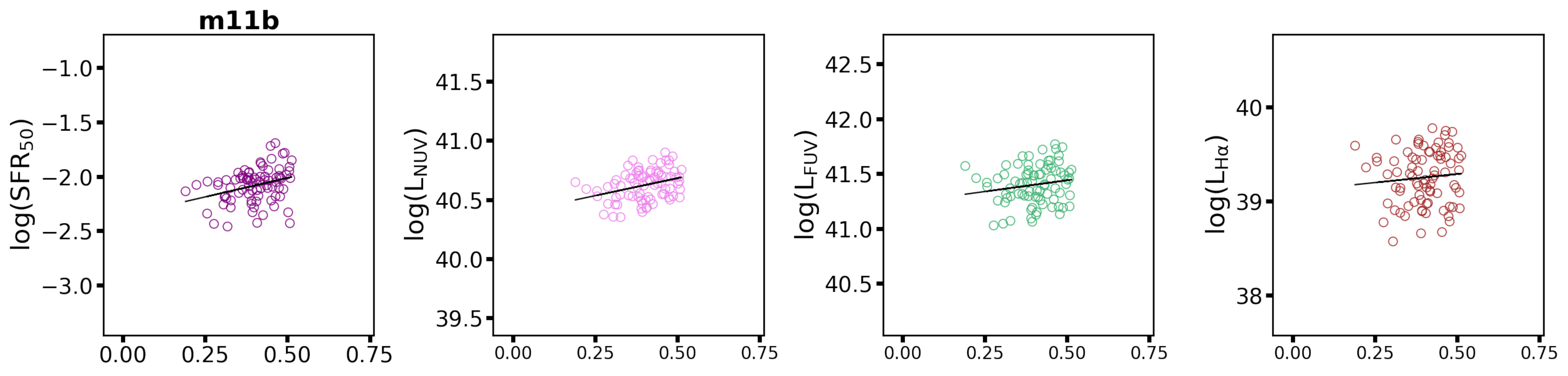}
\includegraphics[width=1\linewidth]{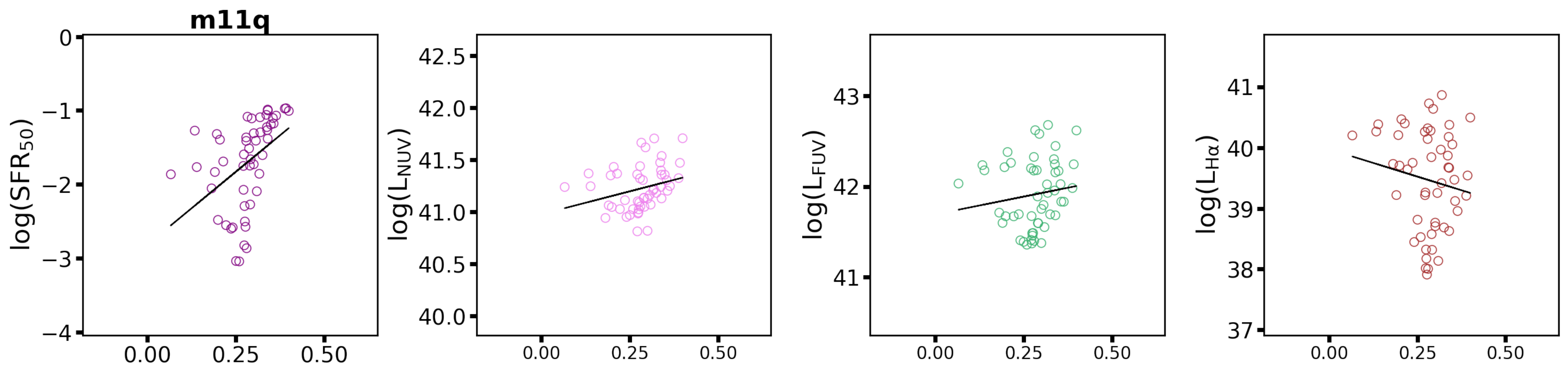}
\includegraphics[width=1\linewidth]{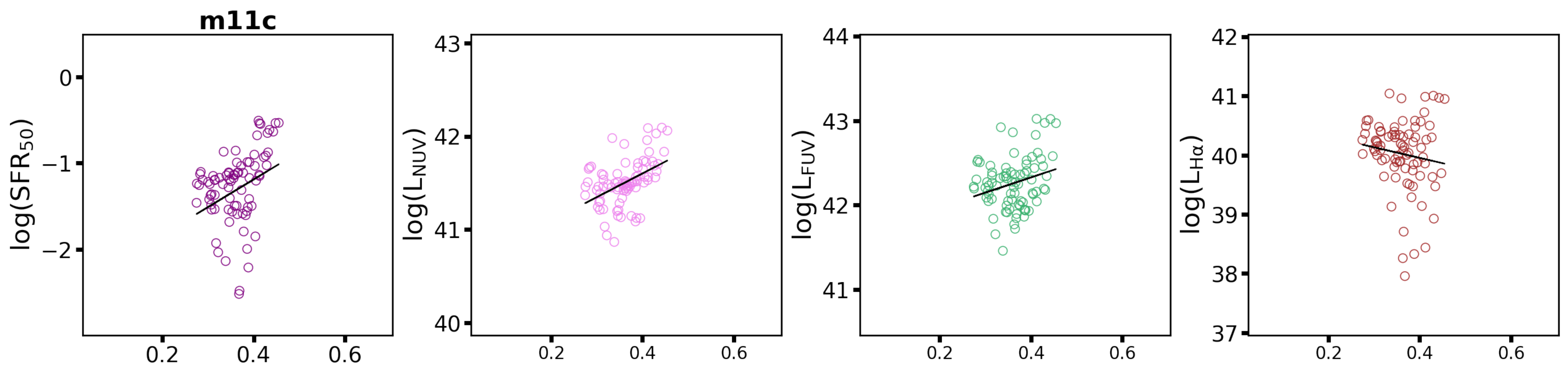}
\end{center}
\end{figure*}

\begin{figure*}
\begin{center}
\includegraphics[width=1\linewidth]{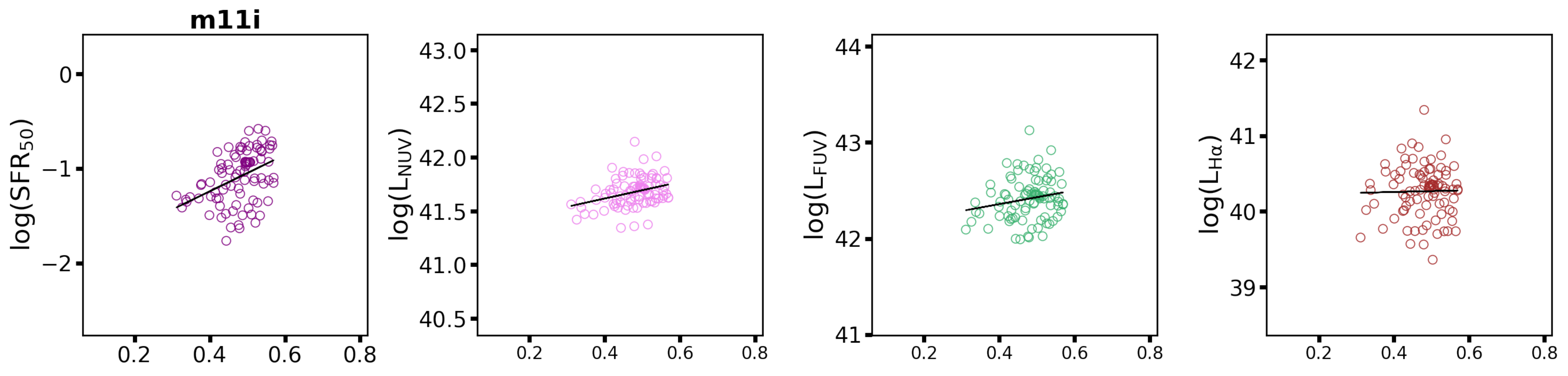}
\includegraphics[width=1\linewidth]{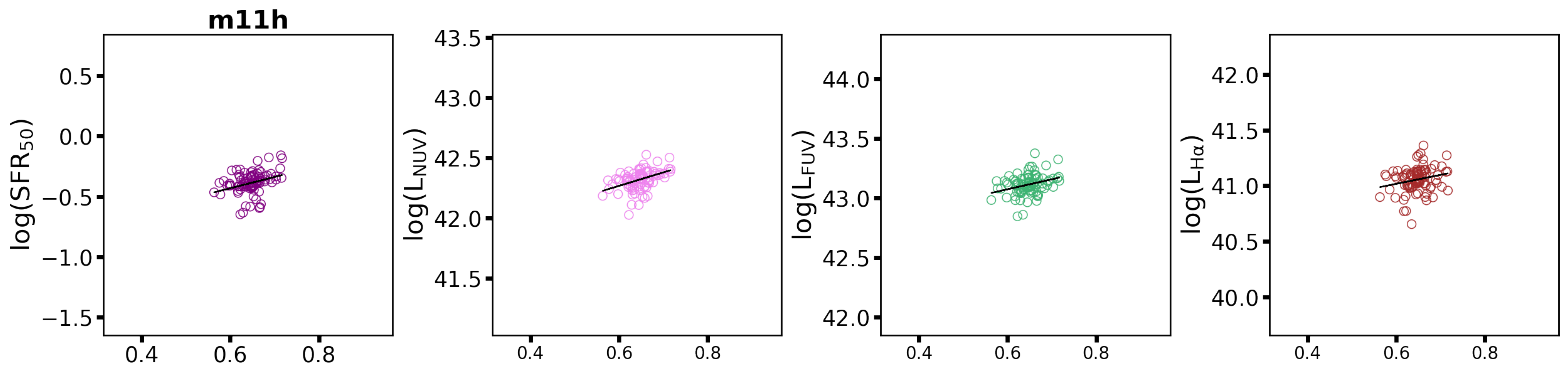}
\includegraphics[width=1\linewidth]{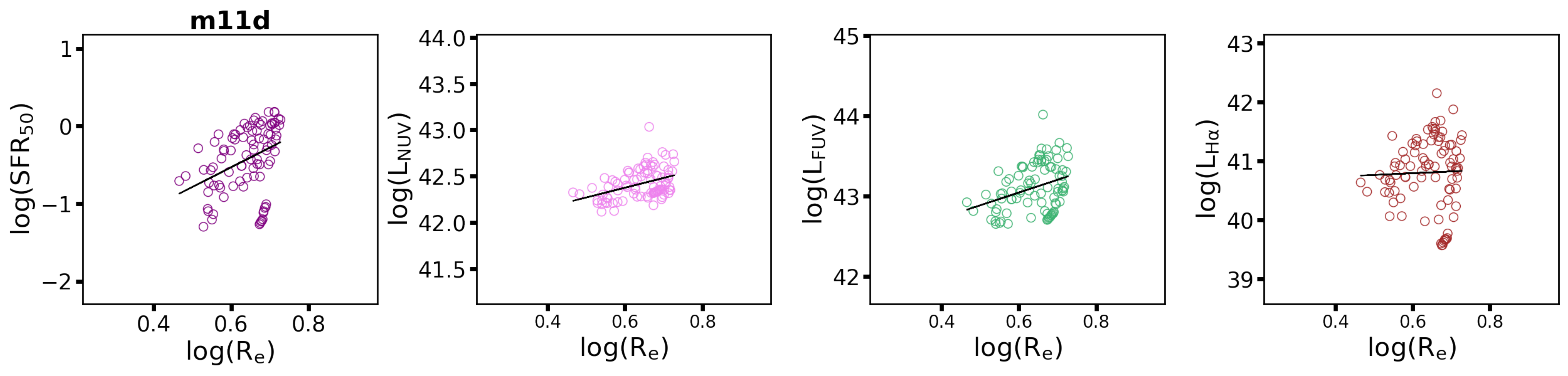}
\caption{Logarithm of different SFR indicators as a function of logarithm of half-light radius (\re) in FIRE-2 galaxies. SFR indicators are sorted based on their correlation with \re, with the highest correlation at left (SFR$_{50}$) and lowest correlation at right (L$_{H\alpha}$). The slope decreases from left to right in most galaxies, which is an emphasis on what has been found in Figures \ref{fig:cross_corr_m11q} and \ref{fig:m11q_timesequences} that the \re\ correlates best with the SFR that is lagged by 50 Myrs.}
\label{fig:re_sfr_fire}
\end{center}
\end{figure*}

\begin{table*}[]
    \centering
    \begin{tabular}{c|c|c|c|c}
        \hline
       Name &  \re-SFR$_{50}$ &  \re-L$_{NUV}$ &  \re-L$_{FUV}$ & \re-L$_{H\alpha}$  \\ &  Slope ($\rho$) & Slope ($\rho$) & Slope ($\rho$) & Slope ($\rho$) \\
       \hline
       \hline
        m10y &  1.28$\pm$1 (0.03)     & 0.62$\pm$1.38 (-0.08)   & 0.12$\pm$1.56 (-0.11)   & -4.22$\pm$2.25 (-0.30)  \\
        m11a &  2.61$\pm$1.05 (0.32)   & 0.59$\pm$0.47 (0.12)   & -0.07$\pm$0.86 (-0.04)  & -3.47$\pm$1.73 (-0.28)  \\
        m11b &  0.70$\pm$0.24 (0.33)   & 0.59$\pm$0.19 (0.32)   & 0.41$\pm$0.27 (0.17)   & 0.37$\pm$0.46 (0.12)  \\
        m11q &  3.94$\pm$1.16 (0.67)  & 0.88$\pm$0.44 (0.38)   & 0.78$\pm$0.80 (0.24)   & -1.80$\pm$1.73 (-0.08)  \\
        m11c &  3.20 $\pm$1.22 (0.39)  & 2.53$\pm$0.55 (0.48)  & 1.80$\pm$0.76 (0.17)   & -1.79$\pm$1.51 (-0.16)  \\
        m11i &  1.93$\pm$0.44 (0.44)  & 0.77$\pm$0.23 (0.33)   & 0.72$\pm$0.38 (0.16)   & 0.11$\pm$0.61 (-0.04) \\
        m11h &  0.93$\pm$0.29 (0.31)  & 1.12$\pm$0.25 (0.45)   & 0.84$\pm$0.27 (0.28)   & 0.81$\pm$0.38 (0.21) \\
        m11d &  2.54$\pm$0.66 (0.44)  & 1.06$\pm$0.25 (0.38)   & 1.59$\pm$0.47 (0.29)   & 0.30$\pm$1.00 (0.03) \\
        
        \hline
    \end{tabular}
    \caption{Slopes and the Spearmann rank correlation ($\rho$) of the log(\re)-log(SFR) relation for different SFR indicators in each FIRE-2 galaxies.}
    \label{table:re_sfr_fire}
\end{table*}

\section{Discussion}
\label{sec:discussion}

\subsection{Previous studies}

\cite{patel_2018} have tested the effect of bursty star formation on the galaxy size fluctuation for a sample of intermediate-mass galaxies (IMGs; $10^9-10^{9.5}$ M$_{\odot}$) at a redshift range of 0.3-0.4 in the Hubble Space Telescope (HST) I$_{814}$ Cosmological Evolution Survey (COSMOS) footprint. They measured the SFRs based on combined UV and IR luminosities and looked at the \re-sSFR relation in their sample. There are some differences between their analyses and ours. The key difference is that \cite{patel_2018} did not subtract off the mass dependency from their \re\ measurements. Thus it becomes impossible to infer whether a large \re\ is because of an expansion due to burstiness or because of an intrinsically massive large galaxy. Despite that, they found that IMGs with higher sSFRs are the most extended with median sizes of \re\ $\sim$ 2.8-3.4 kpc, and IMGs with lower sSFRs are a factor of $\sim$ 2-3 more compact with median sizes of \re\ $\sim$ 0.9-1.3 kpc. This is consistent with our observed log (\normre)-log (sSFR$_{FUV}$) correlation for our $10^7-10^{7.85}$ M$_{\odot}$ local galaxies. They concluded that their observed \re-sSFR correlation is in tension with what simulations predict, i.e., the size should decrease when star formation is in a burst and increase when star formation is quenched. However, we find that both the FUV luminosity and the size of the galaxy trace star formation change on different timescales and the UV luminosity should not be mistaken for the instantaneous SFR. 

As we discussed earlier in Section \ref{sec:Intro}, bursty star formation is predicted to drive fluctuations for both gas and star particles in low mass galaxies in the form of gas emission line broadening and galaxy size fluctuations, respectively \citep{Elbadry_2016}. In this paper, we have tested the effect of burstiness on the galaxy size fluctuations and found that the size of the galaxy is coupled with star formation that is delayed by 50 Myr. Furthermore, there have been observational studies on the star formation histories and its relationship with neutral hydrogen (HI) gas velocities of a sample of local dwarf galaxies \citep{Stilp_2013}. \cite{Stilp_2013} found that HI line broadening, or equivalently HI energy surface density, is strongly correlated only with star formation that occurred 30-40 Myr ago. They argue that the coupling between star formation and the neutral interstellar medium is strongest on this timescale, due either to an intrinsic delay between the release of the peak energy from star formation or to the coherent effects of many supernova explosions during this interval. The timescale found by \cite{Stilp_2013} is very similar to the one found in this work ($\sim$ 50 Myr), and these two findings together, support the theoretical predictions that bursty star formation impacts both stellar and gas kinematics in low mass systems.

\subsection{The role of galaxy formation histories and mergers in \normre-SFR relation}
 
We caution that the fact that bursty star formation can drive size fluctuations does not mean that the size fluctuations are solely driven by burstiness and in reality there might be other factors that might contribute to the size fluctuations at each stellar mass.

We further tested to see whether the relationship between the mass-subtracted sizes (\normre) and the SFRs seen in Figure \ref{fig:lvl_re_sfh} are resulted by other factors than the burstiness, for instance, the galaxy formation histories or the galaxy-galaxy interactions. Galaxies that have formed earlier, indicate small \re\ and tend to lack late-time star formations which reflects as low SFRs in  both H$\alpha$- and UV-derived SFRs. Therefore, there might be a number of early-type galaxies that contribute to our observed \normre\ -SFR trend in Figure \ref{fig:lvl_re_sfh}. Since the H$\alpha$ and UV luminosities of these galaxies are equally low, their L$_{H\alpha}$/L$_{UV}$ approach the equilibrium value of 10$^{-2.12}$ \citep{Emami_2019}. However, a plot of log (L$_{H\alpha}$/L$_{UV}$) of the LVL sample \citep[][Figure 4]{Emami_2019} shows that the low mass bins, i.e. masses below 10$^8$ M$_{\odot}$, span a large range of L$_{H\alpha}$/L$_{UV}$. This implies that galaxies in our LVL sample undergo recent star formations and the \normre\ -SFR trend that we see in our observed sample cannot be driven by the early-type dwarfs. 
Furthermore, galaxies that are undergoing merger processes indicate an increase in their SFRs as well as an increase in their half-light radii due to their tidal tails. Therefore, interacting systems can also result into a rising \normre\ -SFR trend. However we found that none of our galaxies exist in the catalog of known merging local dwarf galaxies \citep{Paudel_2018}. Hence, it is very unlikely that the galaxy formation histories or mergers contribute to the \normre\ -SFR trend that we see in our observed sample. 

\subsection{Implications for high-redshift tests with James Webb Space Telescope}
Our observed findings combined with the findings from the FIRE-2 simulations suggest that the size of a galaxy traces the bursty star formation with a delay of approximately 50 Myrs. Given that bursty star formation is more prominent at high redshifts \citep{Guo_2016, Faisst_2019, caplar_2019}, it is important that we study the effect of burstiness on size fluctuations in a large and complete dwarf galaxy sample at high redshifts. However, at high-redshifts, the photometric aperture in the rest-frame UV and observed optical bands
is severely biased towards compact regions of galaxies where the undergoing bursts occur, and the more extended, old underlying regions are often unresolved \citep[][See also Zick et al. in prep]{Zick_2018, Ma_2018} . This is due to the fact that the extended underlying regions fall below the surface brightness threshold of the current instruments (e.g., Hubble Space Telescope), leading us to underestimate the true size of the galaxies at high redshifts. With the advent of the James Webb Space Telescope and its high surface brightness sensitivity in the near-IR imaging (NIRCam), we can detect the lower surface brightness stellar populations in the galaxies and investigate the morphological effects of bursty star formation on galaxies at early epochs.

\section{Summary}
\label{sec:summary}

We explored the effect of bursty star formation on galaxy sizes, which results from the implementation of stellar feedback onto the dark matter-only simulations. Winds driven by outflows induce fluctuations on the gravitational potential and cause a radial stellar migration in low-mass galaxies. We tested this prediction on samples of both observed local dwarf galaxies and simulated late-time dwarfs across a mass range of $10^7-10^{9.6} M_{\odot}$ where theoretical simulations suggest feedback to be most efficiently altering galaxies kinematics \citep{Elbadry_2016}. For our observed sample, we used H$\alpha$ and far-UV luminosities (L$_{FUV}$) to probe different timescales of the star formation rate as well as the R-band images to measure the half-light radii (\re). We found that the half-light radius correlates well with L$_{FUV}$ while the correlation becomes weaker between \re\ and $L_{H\alpha}$. Furthermore, the \re-L$_{FUV}$ relation is much more significant at masses below $\sim 10^{8} M_{\odot}$ and becomes shallower at higher masses. These findings suggest that the half-light radius changes due to the star formation variations and reacts to this variation within a timescale similar to that of FUV luminosity (i.e., 10 Myrs or longer), but not close to the H$\alpha$ timescale at all, which is a few Myrs. To understand the underlying physics of this size-SFR correlation and determine the approximate timescale over which the size traces the SFR variations, we explored eight low-mass galaxies from the FIRE-2 cosmological simulations at late cosmic times (over the last two Gyrs) that mimic our observed local sample. 
We summarize our findings as follows:

\begin{itemize}

    \item By visual inspections of the half-light radius time sequence and the SFH of galaxies, it is evident that there is a delay from the onset of a burst of star formation to when half-light radius increases. The physical interpretation is that when SF begins rising, gas and stars are highly concentrated and the galaxy is compact. After subsequent supernovae, the resulting outflows drive gas and stars to larger radii and the galaxy expands. This expansion continues until the star formation completely shuts down and then cooled gas and stars migrate back to the center.
    
    \item In order to assess the timescale by which \re\ traces the star formation change, we performed the cross-correlation technique. We found that there is a 50 Myr time lag between the SFH and half-light radius in almost all galaxies, suggesting that the size of the galaxy traces the star formation changes with a 50 Myr delay. This is also evident as a positive slope in the half-light radius-SFR$_{50}$ relation for almost all the FIRE galaxies.
    
    \item Using multiple common SFR observables, including the H$\alpha$, far-UV and near-UV luminosities and cross-correlate these with the half-light radius, we found that the time sequence of near-UV luminosity is more correlated with the half-light radius time sequence, and the correlation decreases for the far-UV and then H$\alpha$ time sequences respectively. This is consistent with our observed results for which half-light radius is better correlated with the far-UV luminosity than the H$\alpha$ luminosity.

    \item The half-light radius-SFR correlation exists in all the FIRE galaxies spanning a mass range of $10^7-10^{9.6} M_{\odot}$. This is in tension with what we found in our observed sample in which the correlation can only be seen in masses below $10^{7.85} M_{\odot}$. This is likely related to the bursty nature of most massive galaxies in the FIRE simulations at low redshifts \citep{Sparre_2017, Emami_2019}.
    
    \item With the high surface brightness sensitivity of the near-IR imaging cameras on the James Webb Space Telescope, we will be able to test the effect of bursty star formation on the size of the high-redshift galaxies while the biases against low surface brightness galaxies are significantly minimized.
    
\end{itemize}

We thank the anonymous referee for providing constructive comments that helped improve the quality of this paper. We also thank T.K. Chan and Amanda Pagul for their helpful discussions.

\bibliography{galacticDynamics.bib}

\section*{appendix}
\label{sec:appendix}

\begin{figure}[hb]
\begin{center}
\includegraphics[width=0.475\linewidth]{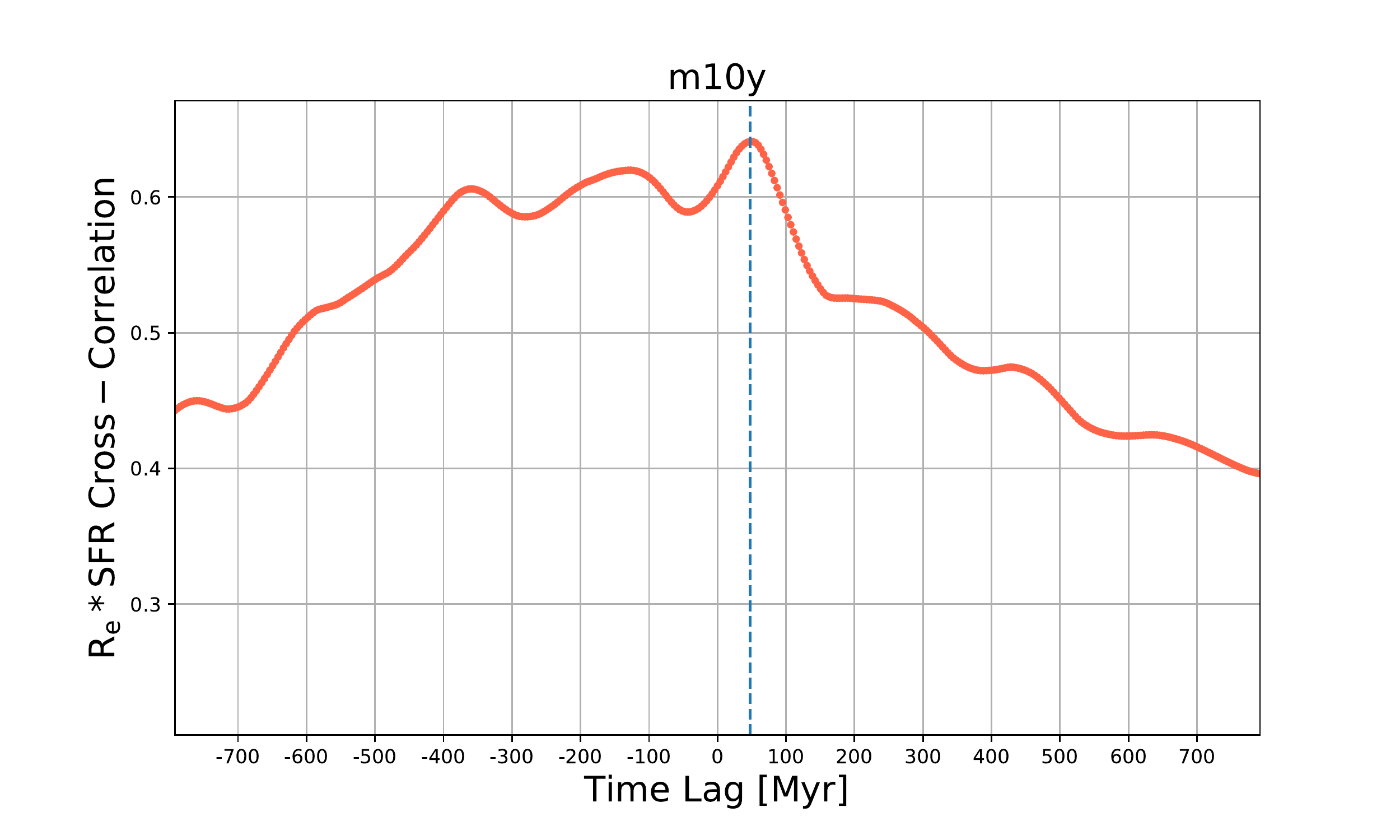}
\hspace{\fill}
\includegraphics[width=0.475\linewidth]{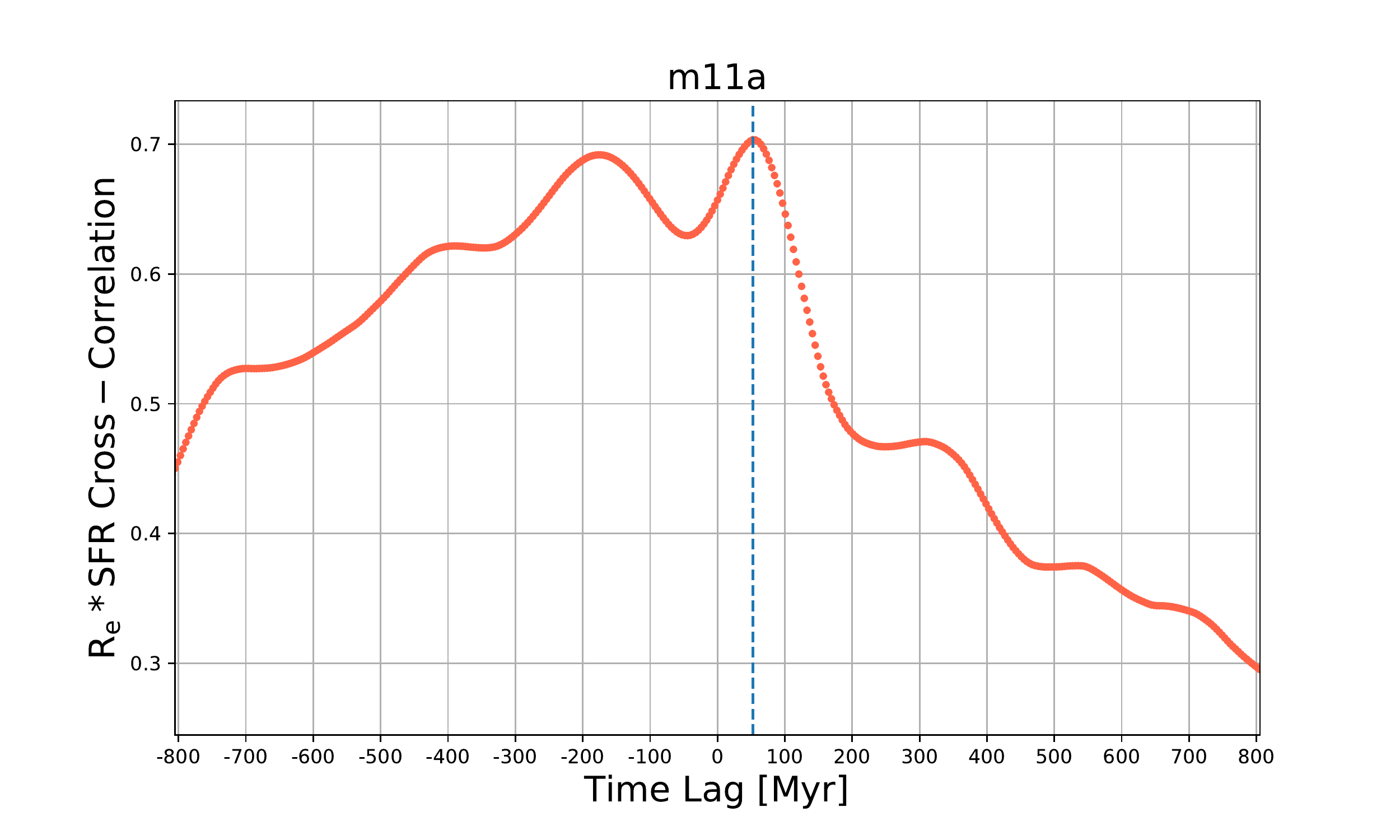}
\hspace{\fill}
\end{center}
\vspace{-0.4in}
\end{figure}

\begin{figure}
\begin{center}
\includegraphics[width=0.475\linewidth]{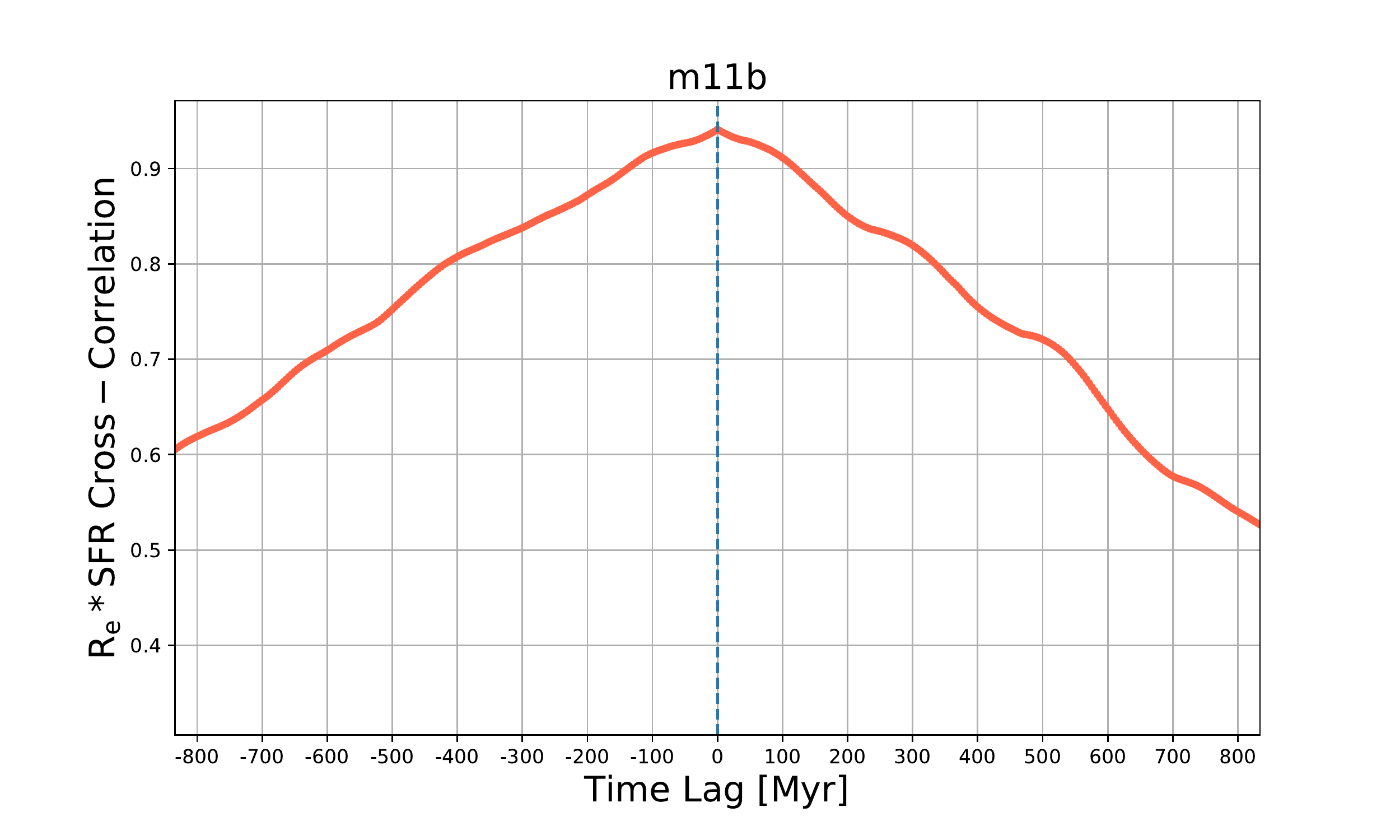}
\hspace{\fill}
\includegraphics[width=0.475\linewidth]{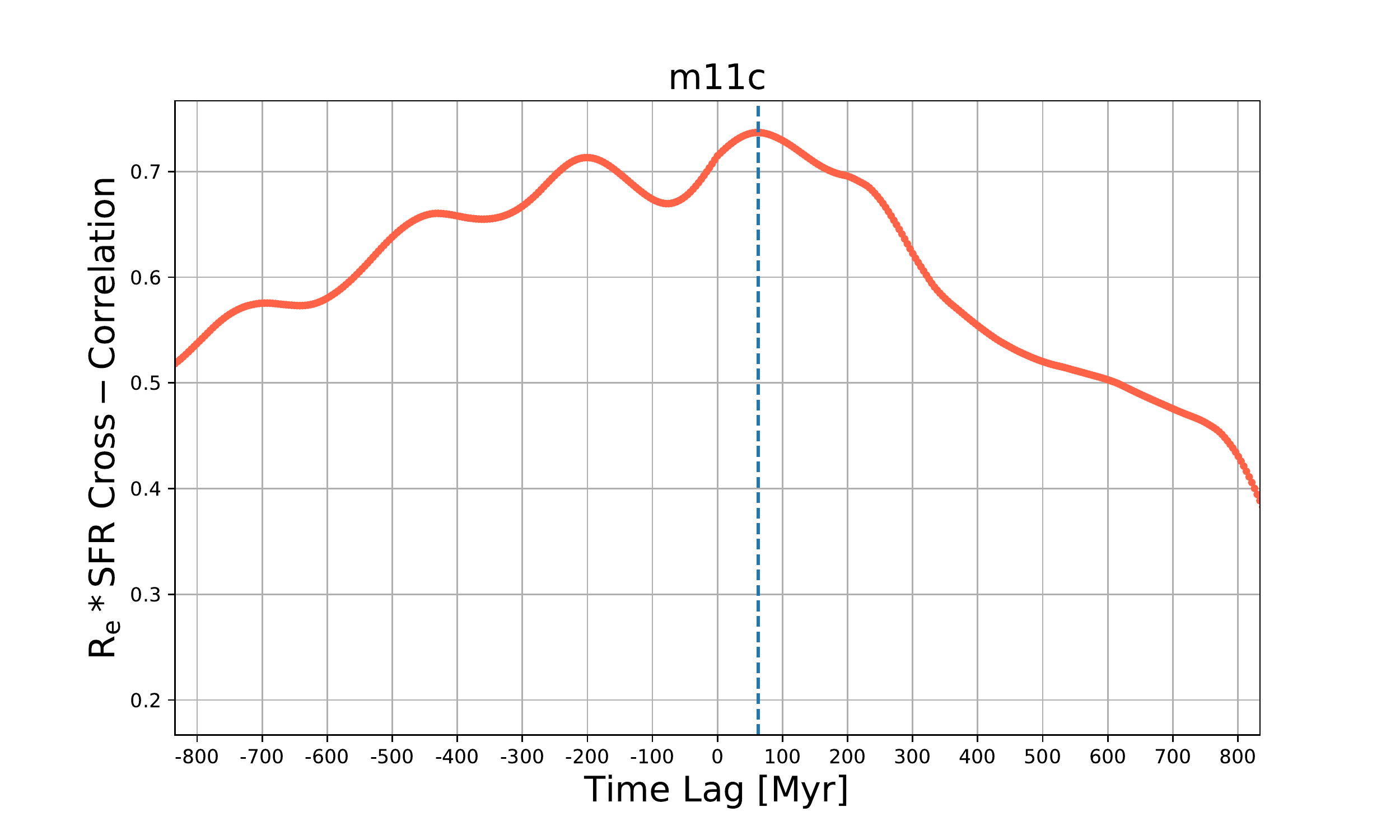}
\hspace{\fill}
\end{center}
\vspace{-0.4in}
\end{figure}

\begin{figure}
\begin{center}
\includegraphics[width=0.475\linewidth]{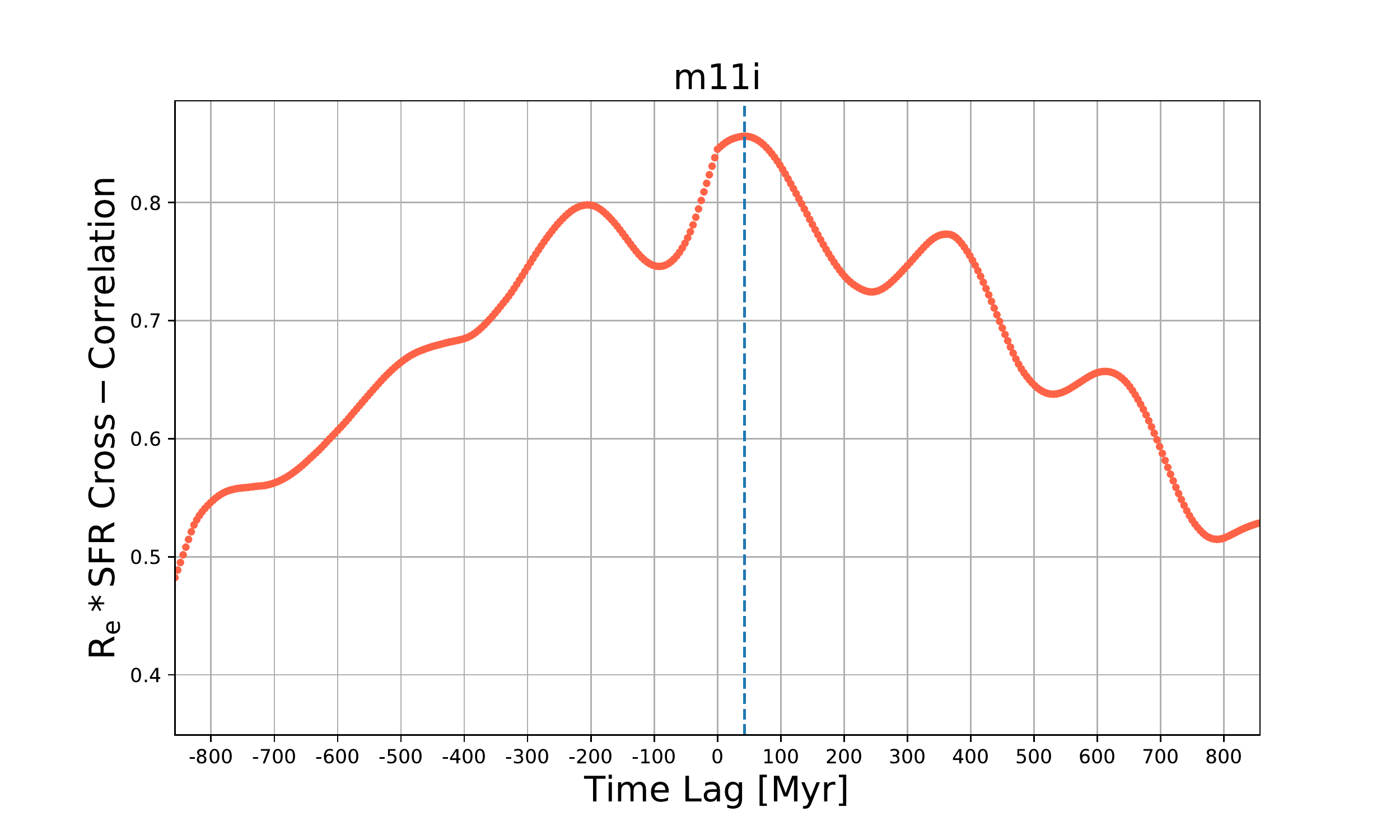}
\hspace{\fill}
\includegraphics[width=0.475\linewidth]{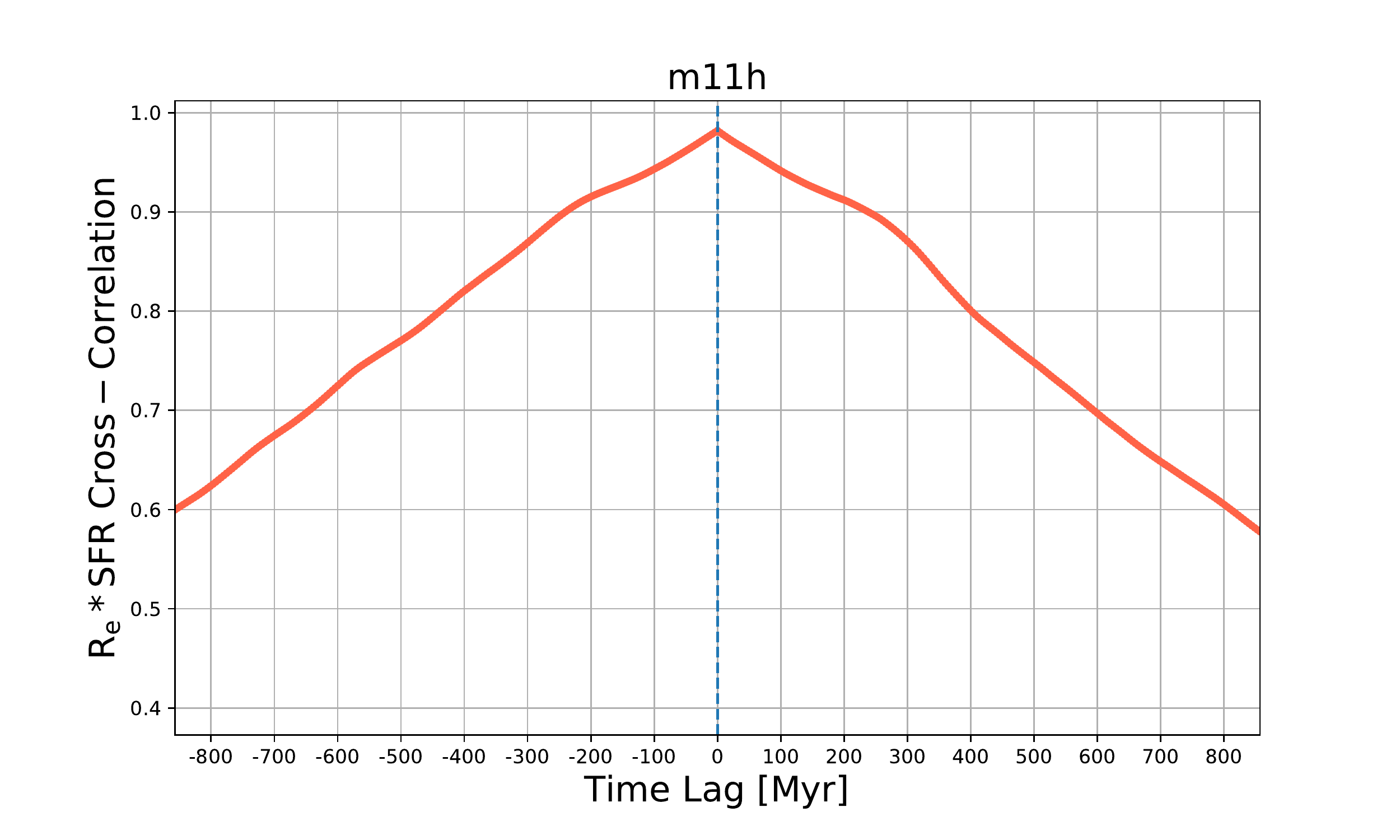}
\hspace{\fill}
\end{center}
\vspace{-0.4in}
\end{figure}

\begin{figure}
\begin{center}
\includegraphics[width=0.475\linewidth]{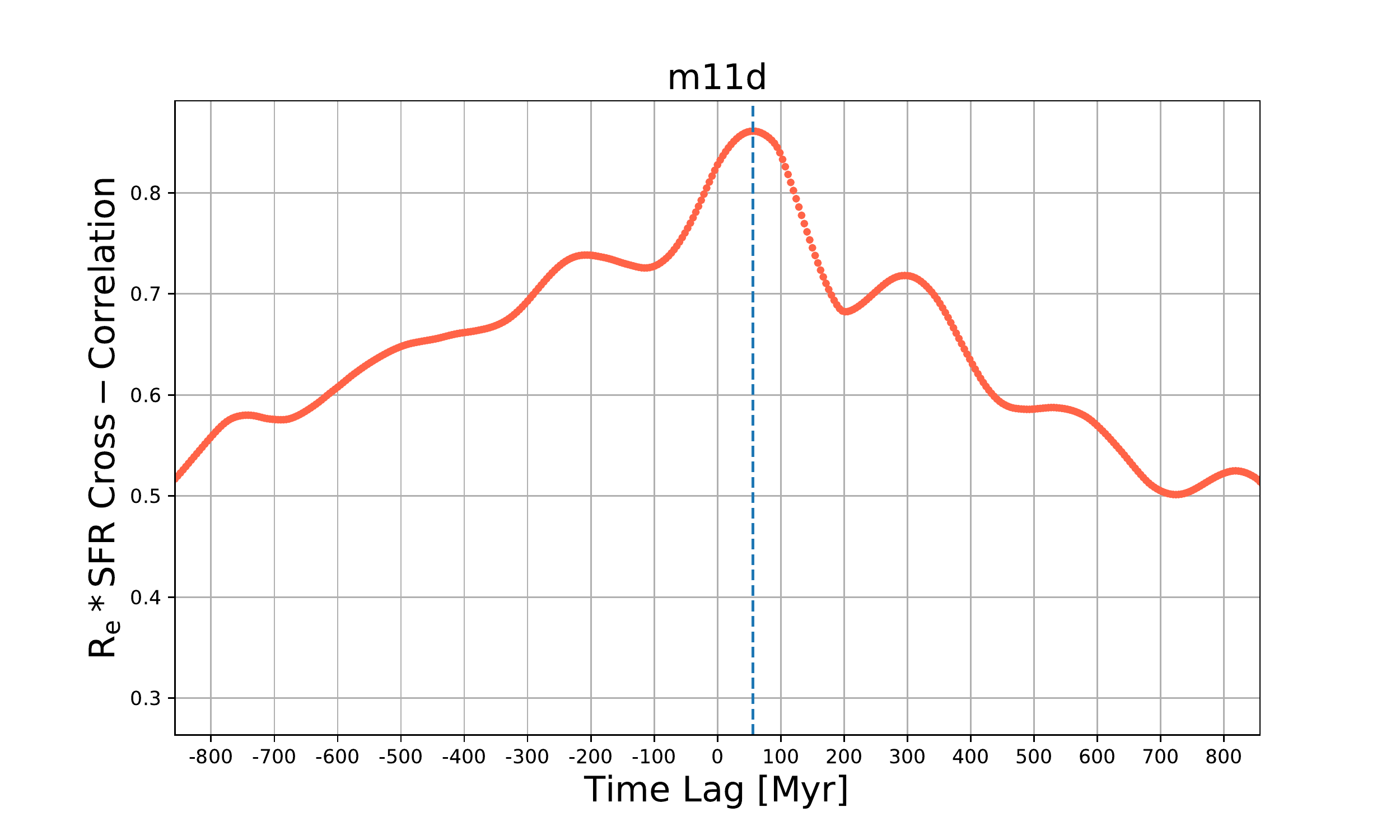}
\hspace{\fill}
\caption{Same as Figure \ref{fig:cross_corr_m11q} for the rest of the FIRE-2 galaxies.}
\label{fig:cross_corr_all}
\end{center}
\end{figure}









\end{document}